\documentclass[pre,eqsecnum,preprint,showpacs]{revtex4}
\usepackage{epsfig}
\usepackage{graphicx}
\usepackage{dcolumn}
\usepackage{bm}
\begin{document}
\title
{Correlations, soliton modes, and
non-Hermitian linear mode\\
transmutation in the 1D noisy Burgers equation}
\author{Hans C. Fogedby}
\email{fogedby@phys.au.dk}
\affiliation
{Institute of Physics and
Astronomy,
University of Aarhus, DK-8000, Aarhus C, Denmark\\
and\\
NORDITA, Blegdamsvej 17, DK-2100, Copenhagen {\O}, Denmark }
\date{\today}
\begin{abstract}
Using the previously developed canonical phase space approach
applied to the noisy Burgers equation in one dimension, we discuss
in detail the growth morphology in terms of nonlinear soliton
modes and superimposed linear modes. We moreover analyze the
non-Hermitian character of the linear mode spectrum and the
associated dynamical pinning and mode transmutation from diffusive
to propagating behavior induced by the solitons. We discuss the
anomalous diffusion of growth modes, switching and pathways,
correlations in the multi-soliton sector, and in detail the
correlations and scaling properties in the two-soliton sector.
\end{abstract}
\maketitle
\section{\label{intro}Introduction}





This is the fourth of a series of papers on the one dimensional
noisy Burgers equation for the slope field of a growing interface.
In paper I \cite{Fogedby98a} we discussed as a prelude the
noiseless Burgers equation \cite{Burgers29,Burgers74} in terms of
its nonlinear soliton or shock wave excitations and performed a
linear stability analysis of the superimposed diffusive mode
spectrum. This analysis provided a heuristic picture of the damped
transient pattern formation. As a continuation of previous work on
the continuum limit of a spin representation of a solid-on-solid
model for a growing interface \cite{Fogedby95}, we applied in
paper II \cite{Fogedby98b} the Martin-Siggia-Rose formalism
\cite{Martin73} in its path integral formulation
\cite{Baussch76,Janssen76,deDominicis78} to the noisy Burgers
equation \cite{Forster76,Forster77} and discussed in the weak
noise limit the growth morphology and scaling properties in terms
of nonlinear soliton excitations with superimposed linear
diffusive modes. In paper III \cite{Fogedby99a} we pursued a
canonical phase space approach based on the weak noise saddle
point approximation to the Martin-Siggia-Rose functional or,
alternatively, the Freidlin-Wentzel symplectic approach to the
Fokker-Planck equation \cite{Freidlin98,Graham89}. This method
provides a dynamical system theory point of view to weak-noise
stochastic processes and yields direct access to the probability
distributions for the noisy Burgers equation; brief accounts of
paper II and III appeared in \cite{Fogedby98c} and
\cite{Fogedby99b}.

Far from equilibrium phenomena are common, including turbulence,
interface and growth problems, chemical reactions, and a host of
other phenomena bordering on biology, sociology and economics.
Unlike equilibrium phenomena the nonequilibrium cases are not very
well understood and constitute a major challenge in modern
statistical physics. Here the Burgers equation provides in many
respects the simplest continuum description of a nonlinear initial
value problem in the noiseless case and an open driven nonlinear
system in the noisy case, exhibiting scaling and pattern
formation.

The noisy Burgers equation for the local slope $u(x,t)$ of a
growing interface analyzed in papers II and II has the form
\begin{eqnarray}
\left(\frac{\partial}{\partial t}-\lambda u\nabla\right)u=
\nu\nabla^2u + \nabla\eta; \label{bur}
\end{eqnarray}
here expressed as manifestly invariant under the slope-dependent
nonlinear Galilei transformation
\begin{eqnarray}
&&x\rightarrow x-\lambda u_0t, \\
&&u\rightarrow u + u_0, \label{gal}
\end{eqnarray}
and is equivalent to the much studied Kardar-Parisi-Zhang (KPZ)
equation \cite{Kardar86,Medina89} for the height $h(x,t)$,
$u=\nabla h$ (in a co-moving frame),
\begin{eqnarray}
\frac{\partial h}{\partial t} = \nu\nabla^2h +
\frac{\lambda}{2}(\nabla h)^2 + \eta. \label{kpz}
\end{eqnarray}
The growth equations (\ref{bur}) and (\ref{kpz}) are driven by
short-range correlated Gaussian white noise $\eta$ determined by the
correlation function
\begin{eqnarray}
\langle\eta\eta\rangle(xt) =\Delta\delta(x)\delta(t),\label{noise}
\end{eqnarray}
characterized by the noise strength $\Delta$. In Eqs. (\ref{bur})
and (\ref{kpz}) the damping constant or viscosity $\nu$ measures
the strength of the linear damping term, whereas $\lambda$ control
the nonlinear growth or mode coupling term.

From the analysis in papers II and III it follows that the
stochastic nonequilibrium problem determined by Eqs. (\ref{bur})
and (\ref{noise}) in the singular weak noise limit
$\Delta\rightarrow 0$ can be replaced by two coupled deterministic
Galilean invariant mean field equations coupling the slope field
$u$ to the canonically conjugate noise field $p$
\begin{eqnarray}
\left(\frac{\partial}{\partial t}-\lambda u\nabla\right)u &&=
\nu\nabla^2 u -\nabla^2p, \label{mf1}
\\
\left(\frac{\partial}{\partial t}-\lambda u\nabla\right)p &&=
-\nu\nabla^2 p. \label{mf2}
\end{eqnarray}
In the path integral analysis in paper II Eqs. (\ref{mf1} -
\ref{mf2}) are saddle point equations for the extremal path in the
classical limit $\Delta\rightarrow 0$; in the canonical phase
space approach in paper III they are classical canonical field
equations determining orbits in an associated phase space. This
doubling of dynamical variables in the deterministic description
was also encountered in the spin model discussed in
\cite{Fogedby95}. The noise variable $\eta$ in Eq. (\ref{bur})
emerges as the canonically conjugate momentum variable $p$
coupling to $u$.

As discussed in papers II and III, see also \cite{Fogedby95}, the
field equations (\ref{mf1}) and (\ref{mf2}) in addition to linear
diffusive modes also support two distinct soliton modes or domain
walls, in the static case of the kink-like form
\begin{eqnarray}
u_s^\mu(x)=\mu u\tanh k_s(x-x_0)~,~~k_s=\lambda u/2\nu.
\label{usol}
\end{eqnarray}
Here $k_s$ sets the inverse length scale, $\mu=\pm$ is a parity
index and the soliton is centered at $x_0$.  For $\mu= + 1$ we
have the right hand soliton which is also a solution of the
noiseless Burgers equation for $\eta=0$ in Eq. (\ref{bur}) or for
$p=0$ in Eq. (\ref{mf1}). For $\mu=-1$ we obtain the noise-induced
left hand soliton, a new solution of the coupled equations. The
associated noise field is $p_s=0$ for the noiseless $\mu=+1$
soliton; for the noisy soliton for $\mu= - 1$ we have
\begin{eqnarray}
p_s = 2\nu u\tanh k_sx, \label{psol}
\end{eqnarray}
modulus a constant.

In the noiseless Burgers equation the transient pattern formation
is described by Galilee-boosted right hand solitons connected by
ramp solutions with superimposed damped linear modes
\cite{Fogedby98a,Saffman68,Jackson90,Whitham74}. In the noiseless
KPZ equation for the height $h$ this pattern corresponds to
smoothed downward cusps connected by parabolic segments with
superimposed linear modes \cite{Medina89,Halpin95}. However, in
the noisy case the doubling of soliton solutions alters the
morphology completely. Here the amplitude-matched Galilee-boosted
right and left hand solitons provide a many body description of a
stationary growing interface. On the soliton gas is superimposed a
gas of linear modes which in the linear Edwards-Wilkinson case
\cite{Edwards82} for $\lambda=0$ becomes the diffusive modes of
the noise-driven diffusion equation.

The canonical phase space approach expounded in paper III moreover
provides a deterministic dynamical system theory description of a
growing interface. With an orbit in canonical phase space from an
initial configuration $u^{\text{i}}$ to a final configuration
$u^{\text{f}}$ traversed in time $T$, determined as a specific
initial-final value solution of the field equations (\ref{mf1})
and (\ref{mf2}), we thus associate the action $S$
\begin{eqnarray}
S(u^{\text{f}},u^{\text{i}},T) = \int_{u^{\text{i}},
0}^{u^{\text{f}}, T}dtdx \left(p\frac{\partial u}{\partial t} -
{\cal H}\right). \label{act}
\end{eqnarray}
More explicitly, the transition probability from an initial
configuration $u^{\text{i}}$ to a final configuration
$u^{\text{f}}$ in time $T$ is determined by
\begin{eqnarray}
&&P(u^{\text{f}}, u^{\text{i}}, T) =
\Omega(T)^{-1}\exp\left[-\frac{1}{\Delta}S(u^{\text{f}},
u^{\text{i}}, T)\right],~~~~ \label{dis}
\\
&&\Omega(T) =\int\prod
du^{\text{f}}\exp\left[-\frac{1}{\Delta}S(u^{\text{f}},
u^{\text{i}}, T)\right],
\end{eqnarray}
where we have introduced the dynamical partition function
$\Omega(T)$, arising from the normalization condition $\int\prod
du^{\text{f}}P(u^{\text{f}}, u^{\text{i}}, T)=1$. Likewise, the
stationary distribution is associated with an infinite-time orbit
from $u^{\text{i}}$ to $u^{\text{f}}$ and is given by
\begin{eqnarray}
P_{\text{st}}(u^{\text{f}}) = \lim_{T\rightarrow\infty}
P(u^{\text{f}}, u^{\text{i}},T),\label{stat}
\end{eqnarray}
and for example the slope correlation (the second moment of $P$) in the
stationary regime by the expression
\begin{eqnarray}
\langle uu\rangle(x,T) = \int\Pi
du^{\text{i}}du^{\text{f}}~u^{\text{f}}(x)u^{\text{i}}(0)
P(u^{\text{f}}, u^{\text{i}}, T)P_{\text{st}}(u^{\text{i}}).
\label{cor}
\end{eqnarray}
The action is a central concept in the weak noise canonical phase
space approach and provides a dynamical weight function and
selection criteria for a dynamical nonequilibrium process in a
similar manner as the energy $E$ in the Boltzmann-Gibbs factor
$P\propto\exp[-\beta E]$, ($\beta$ is the inverse temperature) for
equilibrium processes. The action moreover implies an underlying
principle of least action and the Hamilton entering $S$, yielding
the field equations (\ref{mf1}) and (\ref{mf2}), is given by
\begin{eqnarray}
H  = \int dx~ p(\nu\nabla^2u + \lambda u\nabla u-(1/2)\nabla^2p),
\label{ham}
\end{eqnarray}
where $H = \int dx~ {\cal H}$.

In addition to the conserved Hamiltonian or energy the
translational invariance of $H$ (assuming periodic boundary
conditions for $u$ and for $p$, modulus a constant) implies
conservation of momentum $\Pi$. Moreover, the conserved noise in
Eq. (\ref{bur}), corresponding to the term $\nabla^2p$ in Eq.
(\ref{mf1}), yields the local conservation law $\partial
u/\partial t + \nabla j = 0$, $j=-\nu\nabla u+\nabla
p-(\lambda/2)u^2$, implying the conservation of the integrated
slope field or height offset. The two additional conserved
quantities are thus given by
\begin{eqnarray}
&&\Pi = \int dx~u\nabla p, \label{mom}
\\
&&M = \int dx~ u. \label{intslope}
\end{eqnarray}
We note that the integrated noise field $\tilde P$ is not
conserved for $\lambda\neq 0$. According to Eq. (\ref{mf1})
$\tilde P$ evolves according to, see also \cite{Fogedby95},
\begin{eqnarray}
&&\frac{d\tilde P}{dt} = \lambda\Pi,\label{intnoise}
\\
&&\tilde P = \int dx~p.
\end{eqnarray}

The long time-large distance scaling properties of a growing
interface is a fundamental issue which has been addressed
extensively
\cite{Kardar86,Medina89,Hwa91,Frey96,Tauber95,Frey99,Laessig95,
Laessig98a,Laessig00,Praehofer00a,Colaiori01a,Colaiori01b,Fogedby01a}.
For the width $w(L,t)$ of an interface of size $L$ the dynamical
scaling hypothesis \cite{Halpin95,Barabasi95} asserts that
$w=L^{-\xi}\tilde{G}(t/L^z)$ which for the stationary slope
correlations corresponds to the asymptotic scaling form
\begin{eqnarray}
\langle uu\rangle(x,t) = x^{2\zeta-2}\tilde
F\left(\frac{t}{x^z}\right),\label{scal}
\end{eqnarray}
with roughness exponent $\zeta = 1/2$, dynamical exponent $z$, and
universal scaling function $\tilde F(w)$. In one dimension the
scaling exponents for the noisy Burgers equation are known exactly
\cite{Medina89,Halpin95}. The roughness exponent $\zeta = 1/2$
follows from the known stationary distribution, an effective
fluctuation-dissipation theorem, \cite{Huse85}
\begin{eqnarray}
P_{\text{st}}(u)\propto\exp\left[-\frac{\nu}{\Delta}\int
dx~u^2\right], \label{stat2}
\end{eqnarray}
whereas the dynamic exponent $z=3/2$ is a consequence of the scaling
law
\begin{eqnarray}
\zeta + z = 2,\label{law}
\end{eqnarray}
implied by Galilean invariance \cite{Kardar86,Medina89}. It was an
important result of the analysis in papers II and III, see also
\cite{Fogedby95}, that the dynamical exponent $z=3/2$ also enters
in the dispersion law $E\propto\Pi^z$ for the noise-induced left
hand soliton and thus is a feature of the gapless nonlinear
excitations providing the many body description of a growing
interface.

The description of the stochastic nonequilibrium dynamics of a
growing interface can be accessed on two levels: The Langevin
level defined by Eqs. (\ref{bur}) and (\ref{noise}) or the
Fokker-Planck level (or Master equation level for discrete models)
characterized by the Fokker-Planck equation associated with the
Burgers equation,
\begin{eqnarray}
\Delta\frac{\partial P}{\partial t} = -HP.\label{fpe}
\end{eqnarray}
Here the Hamiltonian or Liouvillian $H$ is given by Eq.
(\ref{ham}), with the momentum variable $p$ interpreted as the
functional derivative $p=\Delta\delta/\delta u$, see also
\cite{Fogedby95}.

On the Langevin level the growth problem is defined by a
stochastic nonlinear differential equation. Apart from direct
numerical simulations the standard analytical tool as regards
scaling properties is a perturbative renormalization group scheme
based on an expansion in powers of the nonlinear term in Eq.
(\ref{bur}) or Eq. (\ref{kpz}) \cite{Forster77,Kardar86,Medina89}.
This procedure yields renormalization group equations in an
$\epsilon$-expansion about $d=2$, $\epsilon = d-2$, predicts a
kinetic phase transition above $d=2$ from a smooth
Edwards-Wilkinson phase ($\zeta = (2-d)/2, z=2)$ to a rough
Burgers-KPZ phase with nontrivial exponents. In $d=1$ the scheme
yields (fortuitously) the exact exponents $\zeta = 1/2$ and
$z=3/2$. The limitation of the Langevin description is that it
does not provide a simple physical picture of a growing interface
and that the role of the noise can only be discussed and
interpreted on a qualitative level.

On the Fokker-Planck level the growing interface is determined by
the deterministic evolution equation (\ref{fpe}) driven by the
Hamiltonian (\ref{ham}). The formal structure of Eq. (\ref{fpe})
is equivalent to a functional Schr\"{o}dinger equation in
Wick-rotated imaginary time and allows via the Martin-Siggia-Rose
functional integral for a mapping of the growth problem onto a
non-Hermitian quantum field theory as discussed in paper II, see
also \cite{Fogedby95}. The quantum field formulation, in addition
to also providing an alternative framework for perturbative
dynamical renormalization group theory following for example the
Callen-Symanzik scheme \cite{Zinn-Justin89}, permits two new lines
of approach to the growth problem. Firstly, by a mapping of the
Martin-Siggia-Rose path integral onto a directed  polymer in a
quenched random medium \cite{Medina89,Halpin95} the nonequilibrium
problem becomes equivalent to a disorder problem affording a
different perspective on the growth problem and yielding new
insight. The second line of approach which we adhere to in the
present context is to discuss the non equilibrium problem directly
in terms of field theoretical constructs. The original stochastic
fluctuations on the Langevin level are then interpreted as quantum
fluctuations on the Fokker-Planck level, where the noise strength
$\Delta$ in Eqs. (\ref{fpe}) serves the role of an effective
Planck constant.

In the context of canonical quantization the quantum field theory
or quantum many body theory for the interface is defined by Eq.
(\ref{ham}) with the canonical momentum $p=\Delta\delta/\delta u$
replaced by the momentum operator $\hat{p} = -i\Delta\delta/\delta
u$ in a $u$-diagonal basis obeying the canonical commutation
relation $[\hat{p}(x),\hat{u}(x)] = -i\Delta\delta(x-x')$. In the
Edwards-Wilkinson case for $\lambda =0$ we are dealing with a free
field theory and the elementary excitations or (undressed)
quasi-particles are the linear non-propagating diffusive modes
with quadratic dispersion $\omega=\nu k^2$, yielding according to
spectral properties the dynamic exponent $z=2$, defining the
Edwards-Wilkinson universality class. As discussed in paper II it
is also an easy task to evaluate e.g., the slope correlations
(\ref{cor}) as a purely quantum many body calculation. In the
nonlinear Burgers case for $\lambda\neq 0$ we obtain correction to
the linear mode dispersion law. Moreover, the quasi-classical
analysis for $\Delta\rightarrow 0$ in papers II and III also
identifies a nonlinear soliton excitation with dispersion law
$E\propto\lambda\nu^{-1/2}\Pi^{3/2}$. A detailed analysis of the
non-Hermitian quantum field theory has, however, not yet been
achieved and will be considered elsewhere.

In the present paper we continue our investigation of the noisy
Burgers equation for the nonequilibrium growth of an interface. We
make use of the weak noise canonical phase space approach
developed in paper III and consider the following important
issues: i) The detailed growth morphology based on the
multi-soliton many body description, ii) the non-Hermitian
properties of the superimposed linear mode spectrum and the
phenomenon of dynamical pinning and mode transmutation, and iii)
the correlations in the Edwards-Wilkinson case, the anomalous
diffusion of growth modes and switching and pathways in the
Burgers-KPZ case, correlations in the multi-soliton sector, and
correlations and scaling in the two-soliton sector. With respect
to i) we stress that one of the advantages of the quasi-classical
weak noise phase space approach propounded in papers II and III is
that it provides a many body description of a growing interface in
terms of solitons and linear modes. The Landau quasi-particle
picture emerging on the Fokker-Planck level was discussed
heuristically in paper II. Here we analyze in more detail the time
evolution of a growing interface in terms of its elementary
excitations. Regarding ii) we note that superimposed on the
nonlinear solitons are linear modes obtained by a linear stability
analysis of Eqs. (\ref{mf1}) and (\ref{mf2}) about a soliton mode.
An analysis of the linear mode spectrum was initiated in paper I
and II. In the present paper we complete the analysis also for a
multi-soliton state and demonstrate among other properties that
the linear modes subject to the nonlinear soliton modes undergo a
mode transmutation from diffusive non-propagating behavior in the
absence of solitons to propagating behavior in the soliton case.
Finally, with regard to iii) on the scaling properties of the
slope correlations (\ref{cor}) we provided in paper II only a
heuristic expression for the scaling function $\tilde F$ based on
a general spectral representation. Here we amend this situation
and present an explicit expression for $\tilde F$ within the
two-soliton approximation. For brief accounts of the present work
we refer to \cite{Fogedby01a,Fogedby01b}; moreover, the papers
\cite{Fogedby02a} and \cite{Fogedby02b} present a numerical
analysis of the soliton-bearing mean field equations and a
tutorial review, respectively.

The paper is organized in the following manner. In
Sec.~\ref{growint} we discuss the growing interface in terms of
soliton modes, in Sec.~\ref{fluc} we consider the superimposed
linear mode spectrum and discuss the mode transmutation alluded to
above, in Sec.~\ref{corr} we address the statistical properties,
and consider the anomalous diffusion of growth modes, switching
and pathways, correlations in the multi-soliton sector, and in
detail the correlations in the tractable two-soliton sector. In
Sec.~\ref{sum} we present a summary and a conclusion.
\section{\label{growint}A growing interface}
A growing interface governed by the noisy Burgers equation
(\ref{bur}) is a simple prototype of an intrinsically open and
driven nonequilibrium system. In the noiseless case for $\eta =0$
the interface is damped and the slope field $u$ evolves subject to
a transient pattern formation consisting of propagating and
merging right hand solitons connected by ramp solutions, with
superimposed damped linear modes \cite{Fogedby98a,Woyczynski98}.
The motion is deterministic and non-fluctuational. At long times
the solitons die out on a time scale set by $1/\nu k^2$, where $k$
is the diffusive mode wavenumber. In the noisy case for $\eta\neq
0$ the time evolution and pattern formation change. The noise
balances the damping and drives after a transient period the slope
field into a stationary morphology composed of amplitude-matched
right and left hand solitons with superimposed linear modes. The
noise strength $\Delta$ is an essential parameter changing the
qualitative morphology of an interface; this is reflected
mathematically in Eq. (\ref{dis}) for the transition probability
which has an essential singularity for $\Delta\rightarrow 0$.

The above behavior is illustrated in the linear case where an
explicit solution of Eq. (\ref{bur}) for a wavenumber component
$u_k$, $u_k(t)=\int dx~u(x,t)\exp(-ikx)$, of the slope field
driven by the noise wavenumber component $\eta_k$ is given by
\begin{eqnarray}
u_k(t)= u_k^{\text{i}}e^{-\omega_kt} - \int_0^tdt' ik
e^{-\omega_k(t-t')}\eta_k(t'). \label{ulin}
\end{eqnarray}
Here $\omega_k=\nu k^2$ is the diffusive mode dispersion law and
$u^{\text{i}}_k = u_k(t=0)$ the initial slope value. We notice
that generally $1/\omega_k$ sets the time scale. Initially the
motion is deterministic and governed by the noiseless diffusion
equation; at longer times for $\omega_k t\gg 1$ the noise
gradually picks up the motion as indicated by the kernel
$\exp[-\omega_k(t-t')]$ in Eq. (\ref{ulin}) and $u_k $ begins to
fluctuate and is driven into a stationary noisy state. This
behavior is in accordance with the phase space behavior discussed
in paper III on the Fokker-Planck level. In Fig.~\ref{fig1}  we
have for a particular noise realization depicted the behavior of
$u_k$. We emphasize that the general aspects of the noise-induced
time evolution also holds in the noisy Burgers case here subject
to a soliton-induced pattern formation. The transient regime is
indicated by I, the long time stationary regime by II.

As mentioned above the quantum mechanical interpretation allows a
discussion of the growing interface in terms of a Landau
quasi-particle picture. In the Edwards-Wilkinson or noninteracting
case the relevant quasi-particle is the diffusive mode $u_k$ with
quadratic dispersion law $\omega = \nu k^2$. In the Burgers case
it is a general feature of the Landau quasi-particle picture that
interactions usually give rise to a dressing effect of the free
(bare) quasi-particle, e. g., the inducement of an effective mass.
In the diffusive mode case this corresponds to a dressing of the
damping constant $\nu$ leaving the dynamical exponent $z=2$ in
$\omega = \nu k^2$ unaltered. However, as shown in papers II and
III even for weak $\lambda$ a new quasi-particle emerges, the
nonlinear soliton excitation, with dispersion law
$\omega\propto(\lambda/\nu^{1/2})k^{3/2}$.

From a heuristic point of view we can regard the soliton as a
self-bound state of diffusive modes; in other words, the solitons
condense or nucleate out of the diffusive mode field. We note,
however, that while the formation of localized soliton modes with
superimposed linear modes is a well-known feature of deterministic
evolution equations, e.g, the sine-Gordon equation and the
nonlinear Schr\"{o}dinger equation \cite{Scott99}, the underlying
mechanism of the doubling of soliton modes here is the noise. In
the approach in \cite{Fogedby95} the soliton mode was identified
as a special solution of the classical field equations (\ref{mf1})
and (\ref{mf2}) obtained in the limit $\Delta\rightarrow 0$ from
i) in \cite{Fogedby95} the underlying Heisenberg field equations
pertaining to the quantum description and ii) in papers II and III
from the classical field equations arising from a principle of
least action in the WKB limit of the Fokker-Planck description.
Below we turn to a discussion of the fluctuating interface in
terms of the quantum/classical picture discussed above.
\subsection{The Edwards-Wilkinson case - equilibrium interface}
In the Edwards-Wilkinson case \cite{Edwards82,Barabasi95} the
slope of a fluctuating interface is governed by the driven
conserved diffusion equation
\begin{eqnarray}
\frac{\partial u}{\partial t}=\nu\nabla^2 u+\nabla\eta,
\label{dif}
\end{eqnarray}
which is readily solved both on the Langevin level in Eq.
(\ref{ulin}) and on the Fokker-Planck level. Since the damping
term $\nu\nabla^2u$ in Eq.(\ref{dif}) can be derived from a
thermodynamic free energy $F=(1/2)\int dx~u^2$ the driven
diffusive equation describes an interface in equilibrium at a
temperature $T=\Delta/2\nu$. In accordance with the quantum field
interpretation outlined above, this simple case, however, serves
as an illustration of the quasi-particle representation.
Consequently, we turn to the field equations (\ref{mf1}) and
(\ref{mf2}) in the linear case for $\lambda=0$:
\begin{eqnarray}
\frac{\partial u}{\partial t}&&= \nu\nabla^2 u -\nabla^2p,
\label{lin1}
\\
\frac{\partial p}{\partial t} &&= -\nu\nabla^2 p. \label{lin2}
\end{eqnarray}
For a single wavenumber component Eq. (\ref{dif}) corresponds to a
noise-driven overdamped oscillator with force constant $\omega_k =
\nu k^2$ and the associated canonical field equations (\ref{lin1})
and (\ref{lin2}) were solved and discussed in paper III. We find,
supplementing the analysis in paper III, the solutions
\begin{eqnarray}
u_k(t) &&= \frac{u_k^{\text{f}}\sinh{\omega_kt}
+u_k^{\text{i}}\sinh{\omega_k(T-t)}}{\sinh{\omega_kT}},\label{uk}
\\
p_k(t) &&= \nu e^{\omega_kt}\frac{u_k^{\text{f}} -
u_k^{\text{i}}e^{-\omega_kT}} {\sinh{\omega_kT}} ~, \label{pk}
\end{eqnarray}
for an orbit from $u_k^{\text{i}}$ to $u_k^{\text{f}}$ in time
$T$, $0<t<T$. The noise field $p_k(t)$ is slaved to the motion of
$u_k$ and determined by $u_k^{\text{i}}$, $u_k^{\text{f}}$, and
$T$. During the time evolution it evolves from $p_k^{\text{i}} =
\nu[u_k^{\text{f}} - u_k^{\text{i}}\exp(-\omega_k T)]/
\sinh{\omega_kT}$ to $p_k^{\text{f}} =
\nu[u_k^{\text{f}}\exp(\omega_kT) - u_k^{\text{i}}]/
\sinh{\omega_kT}$.

\begin{figure}
\includegraphics[width=0.6\hsize]
{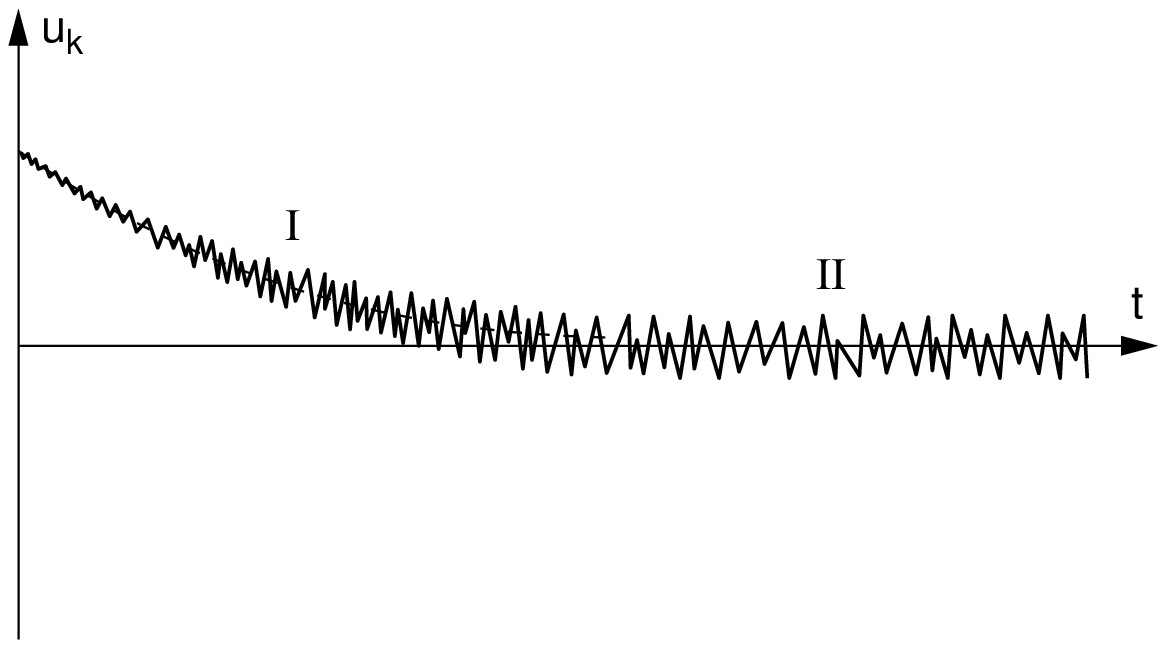}\caption{ We depict the noisy behavior of a
wavenumber component of the slope field, $u_k(t)$, in the
Edwards-Wilkinson case for $\lambda = 0$. After a transient period
given by $1/\omega_k$ the noise on the same time scale gradually
picks up the motion and drives $u_k(t)$ into a stationary noisy
state. The transient regime is denoted by I; the stationary regime
by II.} \label{fig1}
\end{figure}

We note the correspondence between the physical interpretation on
the Langevin level given by Eq. (\ref{ulin}) and on the
Fokker-Planck level characterized by Eqs. (\ref{uk}) and
(\ref{pk}). In the noiseless case for $\eta = 0$, corresponding to
setting  $p_k = 0$, i.e., $u_k^{\text{f}} =
u_k^{\text{i}}\exp(-\omega_kT)$, the slope field is damped
according to $u_k(t) = u_k^{\text{i}}\exp(-\omega_kt)$ over a time
scale $1/\omega_k$. In the presence of noise the growing noise
field $p_k\propto\exp(\omega_k t)$ eventually drives $u_k$, i.e.,
$u_k\propto\exp(\omega_kt)$. Generally, $u_k$ is a linear
combination of a damped part $\exp(-\omega_kt)$ and a growing part
$\exp(\omega_kt)$, analogous to the decomposition of the field in
positive and negative frequency parts in quantum many body theory
\cite{Mahan90}. Here the components are decaying and growing
according to the transient and stationary regimes I and II in
Fig.~\ref{fig1}, respectively.

\begin{figure}
\includegraphics[width=0.6\hsize]
{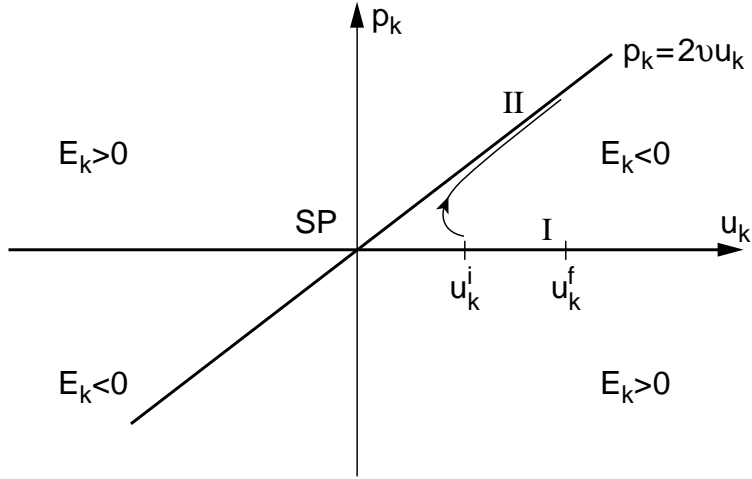} \caption{ Canonical phase space plot in the linear
case for $\lambda = 0$ of a single wavenumber component. The solid
lines indicate the transient submanifold $p_k= 0$ (I) and the
stationary submanifold $p_k = 2\nu u_k$ (II). The stationary
saddle point (SP) is at the origin. For $t\rightarrow\infty$ the
orbit from $u_k^{\text{i}}$ to $u_k^{\text{f}}$ migrates to the
zero-energy manifold. The infinite waiting time at the saddle
point corresponds to ergodic behavior. } \label{fig2}
\end{figure}

The orbit $(u_k,p_k)$ given by Eqs. (\ref{uk}) and (\ref{pk}),
representing the quasi-particles in the classical limit
$\Delta\rightarrow 0$, is confined to a submanifold in phase space
delimited by four global conservation laws: Conservation of energy
$E=H$, conservation of momentum $\Pi$, conservation of area, i.e.,
the integrated slope $M$ or height offset; and here also
conservation of the integrated noise field $\tilde P$ given by
Eqs. (\ref{ham}), (\ref{mom}), (\ref{intslope}),and
(\ref{intnoise}) for $\lambda = 0$, respectively. In wavenumber
space we have
\begin{eqnarray}
E = \int\frac{dk}{2\pi}~E_k=\int\frac{dk}{4\pi}~k^2p_{-k} (p_k -
2\nu u_k), \label{energy}
\end{eqnarray}
and the energy decomposes in contributions  $E_k$ for each
wavenumber mode $k$. Inserting Eqs. (\ref{uk}) and (\ref{pk}) we
obtain specifically
\begin{eqnarray}
E_k = \frac{\omega_k^2}{2}
\frac{|u_k^{\text{f}}|^2+|u_k^{\text{i}}|^2-
2u_k^{\text{f}}u_{-k}^{\text{i}}\cosh\omega_kT}{\sinh^2\omega_kT},
\label{energyk}
\end{eqnarray}
and the energy only depends on the initial and final
configurations $u_k^{\text{i}}$, $u_k^{\text{f}}$ and the time
interval $T$. For fixed $u_k^{\text{i}}$ and $u_k^{\text{f}}$ and
in the long time limit $T\rightarrow\infty$ the energy
$E_k\rightarrow 0$ and the orbit migrates to the zero energy
manifolds: $p_k = 0$, the transient noiseless submanifold, and
$p_k = 2\nu u_k$, the stationary noisy submanifold. The orbit thus
asymptotically passes through the saddle point $(u_k,p_k) = (0,0)$
where the diverging waiting time ensures ergodic behavior. In
Fig.~\ref{fig2} we have depicted the orbits in $(u_k,p_k)$ phase
space.

In a similar manner, the momentum $\Pi$ decomposes according to
\begin{eqnarray}
\Pi =\int\frac{dk}{2\pi}~\Pi_k = \int\frac{dk}{2\pi}~iku_{-k}p_k.
\label{momentum}
\end{eqnarray}
We note that $\Pi$ vanishes on the zero-energy manifolds  $p_k=0$
and $p_k=2\nu u_k$; in the latter case since the integral in Eq.
(\ref{momentum}) becomes a total derivative. For a finite time
orbit insertion of Eqs. (\ref{uk}) and (\ref{pk}) explicitly
yields
\begin{eqnarray}
\Pi_k = \nu\frac{\text{Im}(u_k^{\text{i}}u_k^{\text{f}})}
{\sinh\omega_kT},\label{momentumk}
\end{eqnarray}
in terms of $u_k^{\text{i}}$, $u_k^{\text{f}}$, and $T$; for
$T\rightarrow\infty$ we have $\Pi_k\rightarrow 0$. Likewise for
the integrated slope and noise field, $\tilde P=2\nu M$, we have
\begin{eqnarray}
M=\int dx~u^{\text{i}}(x) = u_{k=0}^{\text{i}}.
\end{eqnarray}
Finally, the action associated with an orbit is obtained from Eq.
(\ref{act}). Inserting the equation of motion Eq. (\ref{lin1}) we
have as an intermediate result $S=(1/2\pi)\int dk S_k$,
$S_k=(1/2)\int dtk^2|p_k|^2$ and using Eq. (\ref{pk}) the action
\begin{eqnarray}
S = \nu\int\frac{dk}{2\pi}\frac{|u_k^{\text{f}}-u_k^{\text{i}}
\exp(-\omega_kT)|^2}{1-\exp(-2\omega_kT)},
\end{eqnarray}
determined by the initial and final configurations
$u_k^{\text{i}}$ and $u_k^{\text{f}}$ and the traversal time $T$.
According to Eq. (\ref{dis}) we subsequently obtain the transition
probability
\begin{eqnarray}
P(u_k^{\text{f}},u_k^{\text{i}},T)\propto
\exp{\left[-\frac{\nu}{\Delta}\int\frac{dk}{2\pi}
\frac{|u_k^{\text{f}}-u_k^{\text{i}}
\exp(-\omega_kT)|^2}{1-\exp(-2\omega_kT)}\right]},\label{dis2}
\end{eqnarray}
a well-known result \cite{Gardiner97,Risken89}). In the limit
$T\rightarrow\infty$ the orbit migrates to transient-stationary
zero-energy manifolds and we arrive at the stationary Gaussian
distribution (\ref{stat2}). This behavior in phase space is
consistent with the qualitative behavior shown in Fig.~\ref{fig1}.

Summarizing, in the linear Edwards-Wilkinson case the conserved
noise-driven stochastic diffusion equation is in the weak noise
limit equivalent to coupled field equations admitting both damped
and growing solutions for the slope field.  The stochastic noise
is replaced by a noise field canonically conjugate to the slope
field. Both damped and growing solutions are required in order to
describe the crossover from the transient regime to the stationary
regime. The wavenumber $k$ is a good quantum number and we can
envisage the fluctuating interface as a gas of growing and damped
diffusive modes according to the decomposition, see also paper II,
\begin{eqnarray}
u(x,t) = \int\frac{dk}{2\pi} [A_ke^{-\omega_kt}e^{ikx} +
B_ke^{\omega_kt}e^{-ikx}]. \label{dec}
\end{eqnarray}
In field theoretical terms $u(x,t)$ is a free field and the
elementary modes are noninteracting. A particular mode lies on the
energy surface $E_k$ and is moreover specified by the conserved
momentum $\Pi_k$. Furthermore, under time evolution the integrated
slope field $M$ and noise field $\tilde P$ are also conserved.
Finally, with the mode is associated an action $S_k$ yielding the
transition probability $P$.

The description based on the field equations (\ref{lin1}) and
(\ref{lin2}) and the associated symplectic structure is basically
classical. Subject to canonical quantization the diffusive modes
become bona fide elementary excitations and a Landau
quasi-particle picture of the interface emerges. The original
noise fluctuations are then interpreted as quantum fluctuations
emerging from the underlying operator algebra. Finally, we note
that subject to a Wick rotation $t\rightarrow it$ the  diffusive
quasi-particles are transformed to dispersive particle-like
quasi-particles with mass $\Delta/2\nu$.
\subsection{The Burgers-KPZ case - nonequilibrium interface}
In the case of a growing interface the situation is more complex.
The general behavior depicted in Fig.~\ref{fig1} still holds in
the sense that the interface evolves from a transient state to a
stochastic stationary state. However, unlike the linear
equilibrium case where the fluctuations in $u$ are extended and
diffusive, the fluctuations in the nonlinear nonequilibrium growth
case include localized propagating modes in order to account for a
growing height profile. This is also evident from e.g., the KPZ
equation in Eq. (\ref{kpz}), where the damping and growth terms
transform differently under time reversal. For $t\rightarrow -t$,
$h\rightarrow -h$, and $\nu\rightarrow -\nu$ the KPZ equation
stays invariant. This is consistent with the fact that in the
decomposition of the irreversible linear modes in growing and
decaying components the damping $\nu$ enters in the combination
$\nu t$, whereas the average nonlinear reversible growth term in
the stationary state $\langle
dh/dt\rangle_{\text{st}}=(\lambda/2)\langle(\nabla h)^2\rangle$ is
invariant. Clearly, the growth term cannot be derived from a
thermodynamic free energy, the term moreover violates the
potential condition \cite{Fogedby80,Stratonovich63,Ma75,Deker75}
and drives the system away from thermal equilibrium into a
stationary kinetic growing state.

The issue on the Fokker-Planck level is again to establish a
quasi-particle picture and to determine orbits in $(u,p)$ phase
space from an initial configuration $u^{\text{i}}$ at time $t=0$
to a final configuration $u^{\text{f}}$ at time $t=T$ in order to,
via the action associated with the orbit, evaluate the transition
probability $P(u^{\text{f}},u^{\text{i}},T)$. The orbit is in
principle determined as an initial-final value problem, i.e., a
boundary value problem in time, of the mean field equations
(\ref{mf1}) and (\ref{mf2}). Unlike the linear case where we can
expand $u$ on plane wave diffusive modes and thus achieve a
complete analysis, the nonlinear and presumably nonintegrable
character of Eqs. (\ref{mf1}) and (\ref{mf2}) precludes such an
analysis.

\begin{figure}
\includegraphics[width=0.6\hsize]
{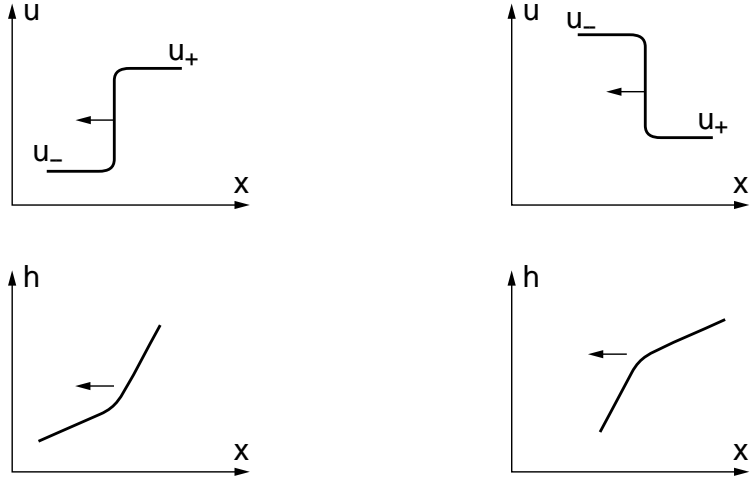} \caption{ We depict the right hand and left hand
moving solitons forming the ``quarks'' in the description of a
growing interface. We have, moreover, shown the associated height
profiles. } \label{fig3}
\end{figure}

Two new features distinguish the field equations (\ref{mf1}) and
(\ref{mf2}) from the linear case: i) the nonlinear coupling
strength $\lambda$ setting together with $\nu$ an intrinsic length
scale $\nu/\lambda$ and ii) the amplitude-dependent Galilean
invariance (\ref{gal}). The new length scale allows for the
possibility of localized nonlinear excitations and the Galilean
symmetry permits the generation of a class of propagating
particle-like excitations from a static solution, i.e., an
excitation at rest. The static excitations are the right and left
hand solitons (\ref{usol}) characterized by the parity index $\mu
= \pm 1$. Boosting a static soliton to the velocity $v$, denoting
the boundary values $u_+$ and $u_-$, and using the Galilean
symmetry in (\ref{gal}) we obtain the fundamental soliton
condition
\begin{eqnarray}
u_+ + u_- = -\frac{2v}{\lambda}. \label{solcon}
\end{eqnarray}
In Fig.~\ref{fig3} we have depicted the fundamental ``quarks'' or
solitons and the associated height profiles.

The right hand soliton for $\mu = +1$ moves on the noiseless
transient submanifold $p=0$ and is a solution of the damped
noiseless Burgers equation, i.e., Eq. (\ref{bur}) for $\eta=0$.
Within the canonical framework the right hand soliton does not
contribute to the dynamics of the interface; according to Eqs.
(\ref{ham}) and (\ref{mom}) with $p=0$ it carries zero energy and
zero momentum. The left hand soliton for $\mu = -1$ is associated
with the noisy or stationary submanifold $p = 2\nu u$ and it
follows from the field equations that it is a solution of the
growing noiseless Burgers equation for $\nu\rightarrow -\nu$. Note
that due to the uneven boundary values $u_+$ and $u_-$ the
solitons are self-sustained dissipative structures driven by
boundary currents as discussed in paper I. The left hand soliton
is endowed with dynamical attributes and carries according to Eqs.
(\ref{ham}), (\ref{mom}), and (\ref{act}) energy, momentum, and
action:
\begin{eqnarray}
E &&= \frac{2}{3}\nu\lambda(u^3_+ -u^3_-),\label{sole}
\\
\Pi &&= \nu(u^2_+ -u^2_-),\label{solp} \\
S &&= \frac{1}{6}\nu\lambda|u_+ -u_-|^3T. \label{solac}
\end{eqnarray}
Since $u_+<u_-$ for a left hand soliton its energy is negative.
Expressing $\Pi$ in the form $\Pi = -(2v\nu/\lambda)(u_+ - u_-)$
using Eq. (\ref{solcon}) it follows that $\Pi$ points in the
direction of $v$. From $\Pi =mv$ we can also associate an
amplitude-dependent mass $m = (2\nu/\lambda)|u_+ - u_-|$ with the
soliton. Finally, the action for a left hand soliton orbit over
time $T$ is positive and Galilean invariant.

In addition to the localized soliton modes the field equations
also support linear modes superimposed on the soliton. These modes
are  obtained by a linear stability analysis of the field
equations and will be discussed in Sec.~\ref{fluc} . In the limit
$\lambda\rightarrow 0$ the soliton modes vanish and the remaining
fluctuations are the diffusive modes of the Edwards-Wilkinson
model.

The field equations (\ref{mf1}) and (\ref{mf2}) are nonlinear and
the soliton solutions do not constitute a complete set in the same
manner as the plane wave decomposition (\ref{dec}) in the
Edwards-Wilkinson case. We shall nevertheless as a working
hypothesis  assume that we can resolve a given initial interface
slope profile $u$ in terms of a gas of right hand and left hand
solitons matched according to the soliton condition
(\ref{solcon}); i.e., with horizontal constant slope segments.
Associated with the soliton representation of $u$ there is also a
soliton representation of the associated noise field $p$. From the
form of the field equations it follows that a multi-soliton
configuration is an approximate solution provided we can control
the overlap contribution $\lambda u_\pm^{\mu_i}\nabla u^{\mu_l}$
arising from the nonlinear term; here $i$ and $l$ denote the i-th
and l-th soliton. Since the soliton width is of order $\nu/\lambda
u$ the overlap only contributes in a region of order $\nu/\lambda
u$ about the soliton center and is small in the inviscid limit
$\nu\rightarrow 0$ and for a dilute soliton gas. Otherwise, we
assume that at least for small $\nu$ we can absorb the correction
term in a linear mode contribution. In summary, we represent a
slope configuration $u$ and the associated noise field
configuration $p$ in terms of a gas of right hand and left hand
solitons matched according to Eq. (\ref{solcon}) with superimposed
linear modes.

This representation of a growing interface is in the spirit of a
Landau quasi-particle picture of an interacting quantum many body
system \cite{Landau80b,Mahan90}. In a heuristic sense we assume
that the interface at a given instant of time is characterized by
a gas of localized soliton modes and extended linear modes. Since
the soliton is not associated with a particle but is a nonlinear
solution of classical field equations the soliton number is not
conserved; in other words, solitons are created and annihilated
subject to collisions.

\begin{figure}
\includegraphics[width=0.6\hsize]
{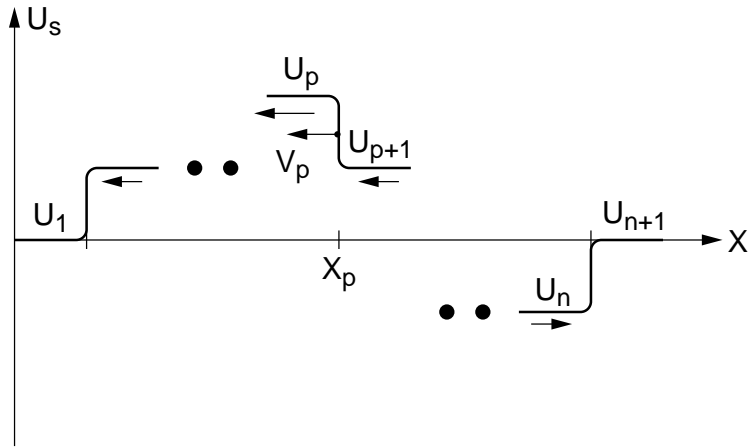} \caption{ We depict an n-soliton slope configuration
of a growing interface. The p-th soliton moves with velocity $v_p
= -(\lambda/2)(u_{p+1}+u_p)$, has boundary values $u_+$ and $u_-$,
and is centered at $x_p$. The arrows on the horizontal
inter-soliton segments indicate the propagation of linear modes.}
\label{fig4}
\end{figure}

Keeping only the solitons, explicit expressions for $u$ and $p$
are easily constructed in terms of the Galilee-boosted soliton
modes (\ref{usol}). Introducing the mean amplitude $k_p$ and
velocity $v_p$ in terms of the boundary value $u_{p+1}$ and $u_p$
of the p-th soliton,
\begin{eqnarray}
&&k_p = \frac{\lambda}{4\nu}(u_{p+1} - u_p),\label{mampl}
\\
&&v_p = -\frac{\lambda}{2}(u_{p+1} + u_p),\label{vel}
\end{eqnarray}
we obtain for an n-soliton representation of the slope field $u_s$
and associated noise field $p_s$
\begin{eqnarray}
&&u_s(x,t) = \frac{2\nu}{\lambda}\sum_{p=1}^n k_p\tanh |k_p|(x
-v_pt-x_p),\label{solconf}
\\
&&p_s(x,t) = \frac{(2\nu)^2}{\lambda}\sum_{p=1,k_p<0}^n k_p\tanh
|k_p|(x -v_pt-x_p).\label{noiseconf}
\end{eqnarray}
The solitons are arranged from left to right according to the
increasing index $p$, $p = 1,2,\cdots n$. The center of the p-th
soliton is at $x_p$ and we have set $u_1=u_{n+1}=0$. In
Fig.~\ref{fig4} we have shown an n-soliton configuration.

\begin{figure}
\includegraphics[width=0.6\hsize]
{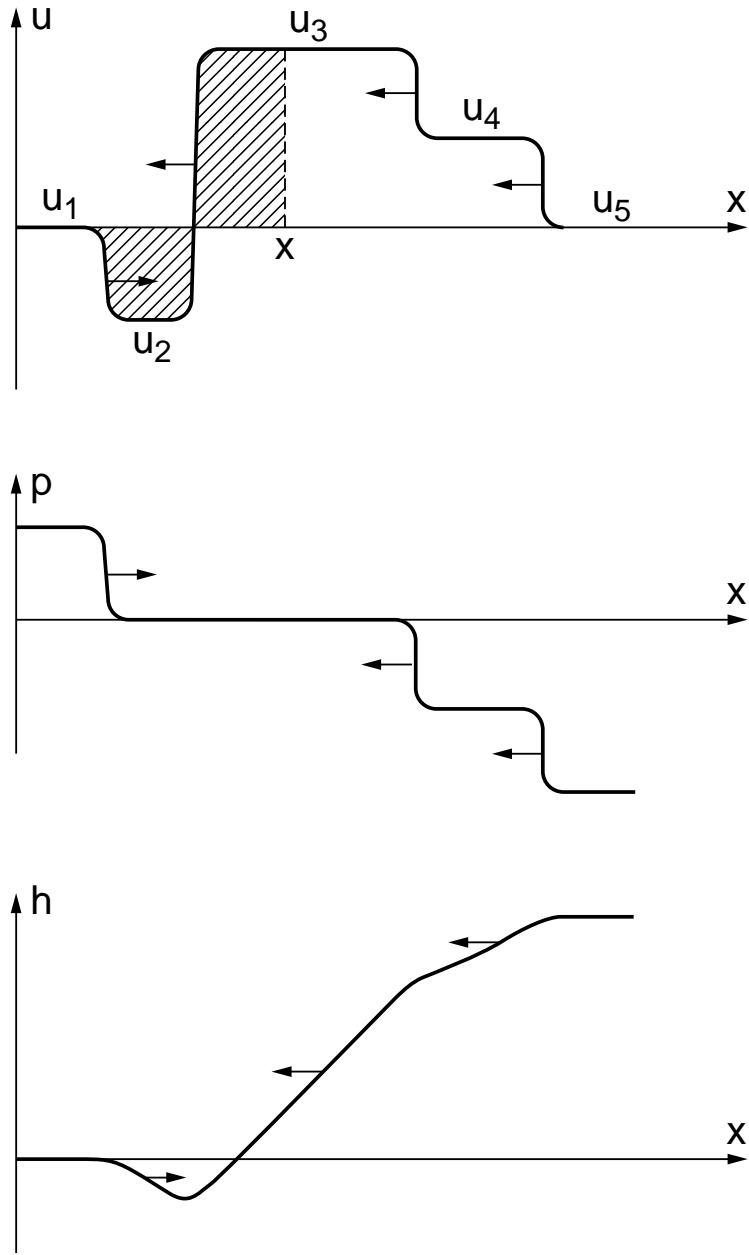} \caption{ We depict the 4-soliton representation of
the slope field $u$, the noise field $p$, and the height field
$h$. The shaded area in $u$, i.e., the integration of $u$ up to
the point $x$ equals the height $h$ at $x$.} \label{fig5}
\end{figure}

The soliton representation in Eq. (\ref{solconf}) of the interface
evolves in time according to the field equations. The motion
corresponds to an orbit in $(u,p)$ phase space lying on the
manifold determined by the conservation of energy $E$, momentum
$\Pi$, and area $M$ according to Eqs. (\ref{ham}), (\ref{mom}),
and (\ref{intslope}). We also observe from Eq. (\ref{intnoise})
that the integrated noise field $\tilde P$, since $\Pi$ is
conserved, develops linearly in time according to
\begin{eqnarray}
\tilde P = \lambda\Pi t + \text{const}.\label{pevol}
\end{eqnarray}
In the soliton representation the contributions to the energy,
momentum, action, and area decompose. Noting that only left hand
solitons for $u_{p+1}<u_{p}$ contribute dynamically we have
applying Eqs. (\ref{sole}), (\ref{solp}), (\ref{solac}) and
(\ref{intslope})
\begin{eqnarray}
&&E = \frac{2}{3}\nu\lambda\sum_{p=1,u_{p+1}<u_p}^n(u_{p+1}^3 -
u_p^3), \label{esol}
\\
&&\Pi = \nu\sum_{p=1,u_{p+1}<u_p}^n(u_{p+1}^2 -
u_p^2),\label{msol}
\\
&&S = \frac{1}{6}\nu\lambda T\sum_{p=1,u_{p+1}<u_p}^n|u_{p+1} -
u_p|^3, \label{actsol}
\\
&&M = \sum_{p=1}^n u_{p+1}(x_{p+1}-x_p). \label{asol}
\end{eqnarray}
Although, as follows from Eqs. (\ref{esol}), (\ref{msol}), and
(\ref{actsol}), the total energy, momentum, and action are
additive quantities (extensive) the soliton gas still represents a
very intricate many body problem. This is due to the soliton
matching condition in Eq. (\ref{solcon}), i.e., the horizontal
segments connecting the solitons, and the dynamical asymmetry
between left hand and right hand solitons. The solitons in the
representation (\ref{solconf}) propagate with in general different
velocities $v_p$ and are thus subject to collisions. Since we only
have at our disposal single soliton solutions of the field
equations patched together to represent a slope configuration at a
particular time instant and not a general solution we have limited
control over soliton-soliton scattering. Clearly, the expressions
(\ref{esol} - \ref{actsol}) only hold in-between soliton
collisions. In particular, the time $T$ entering in the action
(\ref{actsol}) refers to times between collisions, i.e., $T$ is
typically of order $|x_{p+1} - x_p|/v_p$. The working assumption
here is that in-between collisions the soliton (plus linear modes)
representation is valid and that energy, momentum, and area are
conserved during collisions. However, the number of solitons is
not preserved, i.e., solitons are created and annihilated subject
to collisions. Finally, we note that at long times the orbit from
$u^{\text{i}}$ to $u^{\text{f}}$ migrates to the zero-energy
manifold as conjectured in paper III. This implies that the finite
energy solitons at long times are suppressed and that the system
in this limit is  described by diffusive modes yielding the
stationary distribution (\ref{stat2}). For further illustration we
have in Fig.~\ref{fig5} depicted the slope field $u$, height field
$h$, and noise field $p$ for a 4-soliton configuration. The
solitons are centered at $x_1$, $x_2$, $x_3$, and $x_4$ and
propagate with velocities $v_1 = -(\lambda/2)u_2$, $v_2 =
-(\lambda/2)(u_3 + u_2)$, $v_3 = -(\lambda/2)(u_4 + u_3)$, and
$v_4 = -(\lambda/2)u_2$, where $u_2$, $u_3$, and $u_4$ are the
plateau values. The configuration carries energy
$E=(2/3)\nu\lambda(u_2^3-u_3^3)$, momentum $\Pi=\nu(u_2^2-u_3^2)$,
action $S=(1/6)\nu\lambda T(|u_2|^3+|u_4-u_3|^3+|u_4|^3)$, and
area $M=u_2(x_2-x_1)+u_3(x_3-x_2)+u_4(x_4-x_3)$ at time $t=0$. By
integration we note that the total noise field $\tilde P$ evolves
like $\tilde P = \tilde P_0+\lambda\nu t(u_2^2-u_3^2)$ in
agreement with Eq. (\ref{pevol}).

It is clear that the nonequilibrium growth of the interface is
fundamentally related to the existence of localized propagating
soliton modes. Expressing the KPZ equation (\ref{kpz}) in the form
\begin{eqnarray}
\frac{\partial h}{\partial t} = \nu\nabla^2 h +
\frac{\lambda}{2}u^2+\eta, \label{kpz2}
\end{eqnarray}
the linear damping term is associated with the linear modes,
whereas the nonequilibrium growth term is driven by the solitons.
In the Edwards-Wilkinson case the modes are extended and
diffusive, $u\propto\sum_kf_k(t)\cos(kx+\varphi)$, where $f_k(t)$
is related to Eq. (\ref{uk}), and the height field $h=\int^xudx'$
for a particular k-mode behaves like $h\propto\sin(kx+\varphi)$;
consequently, $\langle h\rangle =0$ and growth is absent. On the
contrary, in the Burgers case the localized soliton modes emerge
and $h(x)=\int^xu(x')dx'$ grows owing to the propagation of
solitons across the system. Averaging Eq. (\ref{kpz2}) and setting
$\nu=0$ we have $\langle h\rangle = (\lambda/2)\langle u^2\rangle
t + \text{const.}$ which is consistent with the passage of a
soliton with amplitude $u$ and velocity $|v|=(\lambda/2)u$ at a
given point $x$. This growth behavior also follows from inspection
of Fig.~\ref{fig5}.

The constant slope and noise field configuration $u=u_0$ and
$p=p_0$ have according to Eqs. (\ref{ham}) and (\ref{mom})
vanishing energy $E$ and momentum $\Pi$ and thus form a continuum
of zero-energy states; note that the energy is not bounded from
below, i.e., the zero-energy states are not ground states. The
right and left hand solitons in Eq. (\ref{usol}) lift the
zero-energy degeneracy and connect a constant $u_-$ configuration
to a constant $u_+$ configuration. Unlike the solitons in the
$\varphi^4$ theory or sine-Gordon theory
\cite{Rajaraman87,Scott99}, connecting two degenerate ground
states $\pm\varphi_0$ or degenerate ground states $\varphi_0=\pi
p$, where $p$ is an integer, respectively, with massive gapful
excitations, the Burgers solitons are gapless modes forming a
continuum of states with energy $E\propto(u_+^3-u_-^3)$ and
momentum $\Pi\propto(u_+^2-u_-^2)$.

In the discussion of a growing interface in terms of its slope
field we must introduce appropriate boundary conditions in order
to describe the physical growth state. In the instantaneous
configuration in Figs.~\ref{fig4} and \ref{fig5} we chose for
convenience vanishing slope at the boundaries. However, owing to
the soliton propagation this boundary condition cannot be
maintained in the course of time as the solitons cross the
boundaries of the system and it is more appropriate to assume
periodic boundary conditions for the slope field, i.e., $u(x) =
u(x+L)$ at all times, where $L$ is the size of the system. Note
that periodic boundary conditions for the slope field in general
does not imply periodic boundary conditions for the associated
height field $h$, the integrated slope field. We have from
$h(x)=\int^xu(x')dx'$, $h(x)=h(x+L)+M$, where $M$ is the area
under $u$, and only in the case of  zero-area slope configurations
does $h$ also satisfy periodic boundary conditions, corresponding
to vanishing height offset at the boundaries.

Whereas periodic boundary conditions for the slope field are
consistent with the extended diffusive modes in the
Edwards-Wilkinson case, i.e., the free fields, the elementary
right and left hand Burgers solitons  violate the boundary
conditions since they connect unequal zero-energy states. In this
sense we can regard the solitons as ``quarks'' in the present many
body formulation. A proper elementary excitation or quasi-particle
satisfying periodic boundary conditions is thus composed of two or
more ``quarks'' as illustrated in Figs.~\ref{fig4} and \ref{fig5}.
\subsubsection{The two-soliton configuration}
\begin{figure}
\includegraphics[width=0.6\hsize]
{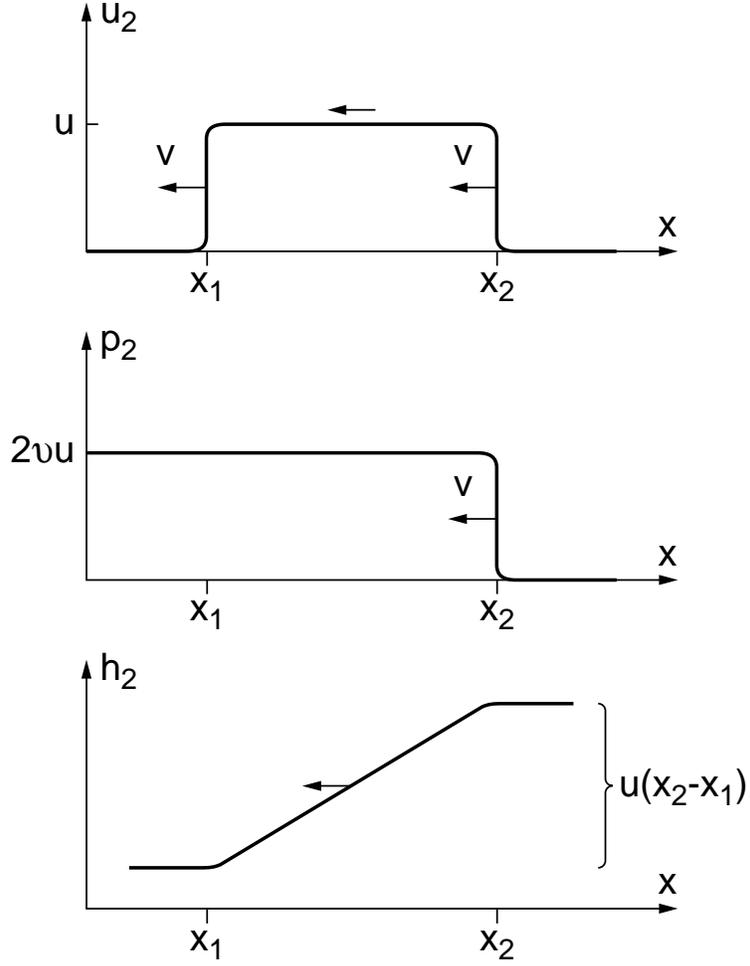} \caption{ We depict the slope field $u_2$, the
associated noise field $p_2$, and the resulting height profile
$h_2$ at time $t=0$ for a two-soliton configuration. The slope
configuration has amplitude $u$, size $\ell = |x_2-x_1|$, and
propagates with velocity $v=-\lambda u/2$. The arrow indicates the
propagation of the superimposed linear mode with phase velocity
$2v$ (discussed in Sec.~\ref{fluc}).} \label{fig6}
\end{figure}

The simplest configuration satisfying periodic boundary condition
is composed of two solitons of opposite parity, i.e., a noisy  and
a noiseless kink. The solitons have the common amplitude $u$, are
centered at $x_1$ and $x_2$ and propagate as a composite entity,
according to Eq. (\ref{solcon}) with velocity $v=-\lambda u/2$.
Specifically, this pair-soliton mode has the form
\begin{eqnarray}
u_2(x,t) = \frac{u}{2}\left[\tanh\frac{k_s}{2}(x-vt-x_1) -
\tanh\frac{k_s}{2}(x-vt-x_2)\right].\label{2u}
\end{eqnarray}
The configuration $u_2$ together with the associated
noise field $p_2$ (for $u>0$)
\begin{eqnarray}
p_2(x,t) = \nu u\left[1-\tanh\frac{k_s}{2}(x-vt-x_2)\right],
\label{2p}
\end{eqnarray}
and the height field $h_2$
\begin{eqnarray}
h_2(x,t) = \frac{u}{k_s}\log \frac {\cosh\frac{k_s}{2}(x-vt-x_1)}
{\cosh\frac{k_s}{2}(x-vt-x_2)} + \text{const.} \label{2h}
\end{eqnarray}
are depicted at the initial time $t=0$ in Fig.~\ref{fig6}

According to Eqs. (\ref{esol} - \ref{asol}) the two-soliton
configuration is endowed with the dynamical attributes:
\begin{eqnarray}
&&E = -\frac{2}{3}\nu\lambda|u|^3, \label{2e}
\\
&&\Pi = \nu\text{sign}(u)u^2, \label{2m}
\\
&&S = \frac{1}{6}\nu\lambda |u|^3T, \label{2ac}
\\
&&M = u(x_2-x_1). \label{2a}
\end{eqnarray}
The two-soliton configuration corresponds to an orbit in $(u,p)$
phase space. Choosing as initial configuration at $t=0$
$(u^{\text{i}}, p^{\text{i}}) = (u_2(x,0),p_2(x,0))$ the final
configuration at $t=T$ is then given by $(u^{\text{f}},
p^{\text{f}}) = (u_2(x,T),p_2(x,T))$. In a finite system of size
$L$ with periodic boundary conditions this is moreover a periodic
orbit with period $L/v$; in the thermodynamic limit
$L\rightarrow\infty$ the period diverges.

By inspection of Fig.~\ref{fig6} it follows that the two-soliton
configuration propagates with a constant profile preserving the
area $M$, i.e., the height offset $2 u(x_2-x_1)$. Subject to
periodic boundary conditions in a system of size $L$ the soliton
pair reappears after a period $L/v$. This motion corresponds to a
simple growth scenario where a layer of thickness $\Delta h =
u(x_2 - x_1)$ is added to $h$ per revolution of the pair. Subject
to this particular soliton mode the interface thus grows with
velocity $(1/2)\lambda u^2(x_2-x_1)/L$. This is consistent with
the averaged form of Eq. (\ref{kpz2}) in the stationary state,
$\langle\partial h/\partial t\rangle = (\lambda/2)\langle
u^2\rangle$, noting the spatial weight $(x_2-x_1)/L$ of the
soliton pair in the average $\langle\partial h/\partial t\rangle$
over the interface. We also remark that the local increase in $h$,
$\Delta h = u\ell$, owing to the passage of a soliton pair of size
$\ell=|x_2-x_1|$ in time $\Delta t=\ell/|v|$, where $|v|=\lambda
u/2$, yields $\Delta h/\Delta t=(\lambda/2)u^2$, again is in
accordance with the averaged KPZ equation. Finally, the integrated
noise field $\tilde P=\int dx p$ decreases linearly with time as
the soliton pair revolves like $\tilde P=\tilde P_0-4\nu u|u|t$ in
agreement with Eq. (\ref{pevol}).

According to $v=-\lambda u/2$ the velocity of the soliton pair is
proportional to the amplitude $u$. Expressing the energy $E$ and
momentum $\Pi$ in terms of $v$ we have
$E=-(16/3)(\nu/\lambda^2)|v|^3$ and $\Pi = 4(\nu/\lambda^2)v|v|$
characterizing the nonlinear excitation. Moreover, eliminating the
velocity we arrive at the dispersion law
\begin{eqnarray}
E = -\frac{4}{3}\frac{\lambda}{\nu^{1/2}}|\Pi|^{3/2}.
\label{soldisp}
\end{eqnarray}
The soliton pair is thus a gapless quasi-particle mode with
exponent $z=3/2$. As discussed in paper II a general spectral
representation for the slope correlations allows us to make
contact with the scaling form in Eq. (\ref{scal}) and identify the
mode exponent $z$ with the dynamic exponent.

Within the present description the statistical weight of the
soliton is determined by Eq. (\ref{stat2}). In the inviscid limit
for small $\nu$ we obtain for a pair of size $\ell$ and amplitude
$u$ the normalized stationary distribution
\begin{eqnarray}
P_{\text{st}}(u,\ell) =
\Omega_{\text{st}}^{-1}(L)\exp\left[-\frac{\nu}{\Delta}u^2\ell\right],
\label{statsol}´
\end{eqnarray}
with normalization factor
\begin{eqnarray}
\Omega_{\text{st}}(L)=2(\pi\Delta/\nu)^{1/2}L^{1/2}.\label{norm1}
\end{eqnarray}
The distribution is parameterized by the amplitude $u$ and the
size $\ell$. The normalization factor or partition function
$\Omega_{\text{st}}$ varies as $L^{1/2}$ and the distribution thus
vanishes in the thermodynamic limit $L\rightarrow\infty$,
characteristic of a localized excitation. The mean size of a
soliton pair is given by
\begin{eqnarray}
\langle\ell\rangle = \int^L_0dl\int du~\ell P_{\text{st}}(u,\ell).
\label{msize}
\end{eqnarray}
Inserting $P_{\text{st}}$ we obtain $\langle\ell\rangle = (1/3)L$,
i.e., the mean size of the pair scales with the system size. This
behavior is characteristic of a spatially extended or loosely
bound excitation and we can envisage the soliton pair as a
``string excitation'' connecting right and left hand solitons (the
fundamental ``quarks'').

In a similar manner we can determine the transition probability
associated with a soliton pair using Eq. (\ref{dis}). Inserting
$S$ from Eq. (\ref{2ac}) and normalizing we obtain
\begin{eqnarray}
P_{\text{sol}}(u,T) = \Omega_{\text{sol}}^{-1}(T)
\exp\left[-\frac{\nu\lambda}{6\Delta}|u|^3T\right],
\label{transol}
\end{eqnarray}
where the normalization factor or dynamic partition function is
given by
\begin{eqnarray}
\Omega_{\text{sol}}(T)=\frac{2}{3}\Gamma\left(\frac{1}{3}\right)
\left[\frac{6\Delta}{\nu\lambda}\frac{1}{T}\right]^{1/3}.
\label{norm2}
\end{eqnarray}
Here the Gamma function $\Gamma(z)=\int_0\exp(-t)t^{z-1}dt$ arises from
the normalization of $P_{\text{sol}}$; $\Gamma(1/3)=2.68174$.

Before turning to the linear fluctuation spectrum in the next
section we wish to briefly review a recent numerical study of the
field equations (\ref{mf1}) and (\ref{mf2}) \cite{Fogedby02a}. The
field equations are of the diffusive-advective type with the
characteristic feature that the equation for $p$ admits
exponentially growing solutions due to the negative diffusion
coefficient thus rendering direct forward integration in time
numerically unfeasible. In order to resolve this instability
problem we developed a ``time loop'' integration procedure based
on integrating the equation for $u$ forward in time followed by an
integration backward in time of the equation for $p$. This
numerical scheme thus requires an assignment of both initial and
final (u,p) configurations and therefore mainly served as a tool
to check whether a certain assignment actually constitutes a
solution.

We investigated numerically three propagating soliton
configurations: i) a propagating soliton pair, ii) two symmetrical
solitons colliding with a static soliton, and iii) the collison of
two symmetrical soliton pairs. Referring to \cite{Fogedby02a} for
details we summarize our findings below. We found in case i) that
the pair-soliton in the inviscid limit for small $\nu$ is a long
lived excitation thus justifying the quasi-particle interpretation
above. In case ii) we considered symmetrical solitons propagating
towards and colliding with a static soliton at the center passing
through the static soliton and reemerging with no phase shift or
delay; this specific mode corresponds to filling in a dip and
subsequently nucleating a tip at the same point in the height
field. In case iii) we finally considered two symmetrical soliton
pairs colliding and reappearing subject to a phase shift or delay
where the incoming trailing solitons become the leading outgoing
solitons; this mode correponds to filling in a trough and
subsequent nucleation of a plateau in the height profile. It was
characteristic of the soliton collisions in case ii) and iii) that
the conservation of $E$, $\Pi$, and $M$ was satisfied during
collision, a feature that seems to stabilize the integration.
\section{\label{fluc} Fluctuations - mode transmutation}
The soliton spectrum discussed in the previous section is a
fundamental signature of the nonlinear character of the Burgers
equation and at the same time essential in accounting for the
growth aspects of an interface. In the present section we address
the fluctuation spectrum  or linear mode spectrum superimposed on
the soliton gas. In the linear Edwards-Wilkinson case discussed in
Sec.~\ref{growint} the fluctuations exhaust the mode spectrum and
have a diffusive character. In the nonlinear Burgers-KPZ case the
fluctuations are superimposed on the soliton modes and become
propagating. The fluctuation spectrum was discussed incompletely
in the noiseless Burgers case in paper I and in the noisy case in
paper II. Here we present a detailed analysis. A brief account of
the work is given in \cite{Fogedby01b}; here we complete the
analysis.
\subsection{The noiseless Burgers case}
The noiseless Burgers equation is inferred from Eq. (\ref{bur}) by
setting  $\eta=0$ and also follows from the field equation
(\ref{mf1}) on the $p=0$ submanifold,
\begin{eqnarray}
\left(\frac{\partial}{\partial t}-\lambda u\nabla\right)u=
\nu\nabla^2u. \label{nbur}
\end{eqnarray}
This equation exhibits a transient pattern formation composed of
right hand solitons connected by ramps with superimposed linear
modes. A single right hand static soliton mode is given by Eq.
(\ref{usol}) for $\mu=1$, i.e., $u_s=u\tanh k_sx$, $k_s=\lambda
u/2\nu$. The soliton has amplitude $u$ and width $k_s^{-1}$. A
spectrum of moving solitons is then generated by the Galilean
boost (\ref{gal}): $x\rightarrow x-\lambda u_0t$, $u\rightarrow u
- u_0$.

In order to analyze the superimposed linear fluctuations we expand
$u$ about the soliton mode $u_s$, $u=u_s+\delta u$. To linear
order in $\delta u$ we obtain the equation of motion
\begin{eqnarray}
\left(\frac{\partial}{\partial t}-\lambda u_s\nabla\right)\delta
u= \nu\nabla^2\delta u + \lambda(\nabla u_s)\delta u.\label{lin0}
\end{eqnarray}
In the asymptotic limit for large $|x|$ this equation is readily
analyzed. Noting that $u_s\rightarrow u\text{sign}(x)$, $\nabla
u_s\rightarrow 0$, and searching for plane wave solutions of the
form $\delta u\propto\exp(-E_kt)\exp(ikx)$ we identify a spectrum
of complex eigenvalues,
\begin{eqnarray}
E_k=\nu k^2 - i\lambda u k\text{sign}(x), \label{spec}
\end{eqnarray}
showing the non-Hermitian character of Eq. (\ref{lin0}).
Introducing the phase velocity $v=\lambda u$ the imaginary part of
the eigenvalue $E_k$ combines with the plane wave part and yields
the propagating damped wave form
\begin{eqnarray}
\delta u\propto e^{-\nu k^2 t}e^{ik(x+vt\text{sign}(x))}.
\label{fluc1}
\end{eqnarray}
The soliton mode thus gives rise to a {\em mode transmutation} in
the sense that the diffusive mode in the Edwards-Wilkinson case
$\delta u\propto \exp (-\nu k^2t) \exp (ikx)$ is transmuted to a
damped propagating mode in the Burgers case with a phase velocity
$v$ depending on the soliton amplitude $u$. The mode
transmutation, of course, also follows from the Galilean
invariance in Eq. (\ref{gal}) since a shift of the slope field to
the soliton amplitude $u$ corresponds to a transformation to the
moving frame $x\rightarrow x-\lambda ut$. For large positive $x$
the mode propagates to the left, for large negative $x$ the
propagation is to the right, i.e., the mode propagates towards the
soliton center which thus acts like a sink. The phenomena of mode
transmutation has also been noted by Sch\"{u}tz \cite{Schuetz00}
in the case of the asymmetric exclusion model, a lattice version
of the noisy Burgers equation, in the context of analyzing the
shocks, corresponding to the solitons in the present context.

As noted in paper I the analysis of Eq. (\ref{lin0}) can be
extended to the whole axis by introducing a nonuniform gauge
function $g$ determined by the soliton profile in Eq.
(\ref{usol}),
\begin{eqnarray}
g(x) = k_s\tanh k_sx~~,~~k_s=\lambda u/2\nu. \label{gf}
\end{eqnarray}
By means of $g$ we can express Eq. (\ref{lin0}) in the
Schr\"{o}dinger-like form
\begin{eqnarray}
-\frac{\partial\delta u}{\partial t} = D(g)\delta u, \label{eve}
\end{eqnarray}
where the operator $D(g)$ in the quantum mechanical analogue is
given by the Hamiltonian
\begin{eqnarray}
D(g) = -\nu(\nabla+g(x))^2 + \nu
k_s^2\left[1-\frac{2}{\cosh^2k_sx}\right]. \label{sham}
\end{eqnarray}
This equation of motion describes in imaginary Wick-rotated time a
particle moving in the potential $-2/\cosh^2k_sx$ subject to the
imaginary gauge field $ig$. Absorbing the gauge field by means of
the gauge transformation
\begin{eqnarray}
U(g) = \exp\left[-\int g(x)~dx\right]=\cosh^{-1}k_sx, \label{gt}
\end{eqnarray}
and using the identity $(\nabla+g)^2 = U(g)\nabla^2U(g)^{-1}$, the
non-Hermitian equation of motion (\ref{eve}) takes the Hermitian
form
\begin{eqnarray}
-\frac{\partial\delta\tilde{u}}{\partial t} = D(0)\delta\tilde{u},
\label{heve}
\end{eqnarray}
where
\begin{eqnarray}
\delta\tilde{u} = U(g)^{-1}\delta u.\label{trans}
\end{eqnarray}
We also observe that the gauge transformation in Eq. (\ref{gt})
has the same form as the Cole-Hopf transformation
\cite{Cole51,Hopf50} applied to the static soliton solution $u_s$,
see also papers I-III.

The presence of the gauge function changes the spatial behavior of the
eigenmodes and thus their normalizability. The imposition of physical
constraints such as spatially localized modes or asymptotic plane wave
modes obeying periodic boundary conditions consequently give rise to a
complex eigenvalue spectrum. Since we can ``gauge'' the mode problem in
Eq. (\ref{eve}) to the exactly solvable Schr\"{o}dinger problem defined
by Eq. (\ref{heve}) we are able to complete the analysis.

Setting $\delta\tilde{u}\propto\exp(-\Omega t)$, where $\Omega$ is
the frequency eigenvalue, the spectrum of the ensuing eigenvalue
problem,
\begin{eqnarray}
D(0)\delta\tilde{u} = \Omega\delta\tilde{u},\label{eve2}
\end{eqnarray}
associated with $D(0)$ can be analyzed analytically
\cite{Fogedby85,Landau59c}. It is composed of a localized zero-
frequency mode and a band of phase-shifted extended scattering
modes with eigenvalue
\begin{eqnarray}
\Omega_k = \nu(k^2+k^2_s).\label{ev}
\end{eqnarray}
The eigenmodes have the form
\begin{eqnarray}
&&\delta\tilde{u}\propto\frac{1}{\cosh
k_sx},~~~~~~~~~~~~~~~~\Omega_0=0, \label{lm}
\\
&&\delta\tilde{u}\propto\exp(ikx)s_k(x), ~~~~~~~~~\Omega_k =
\nu(k^2+k^2_s), \label{sm}
\\
&&s_k(x) = \frac{k+ik_s\tanh k_sx}{k-ik_s}. \label{smt}
\end{eqnarray}
The $x$-dependent s-matrix $s_k(x) = |s_k(x)|\exp[i\delta_k(x)]$
gives rise to a spatial modulation
$|s_k(x)|=[(k^2+k_s^2\tanh^2k_sx)/(k^2+k_s^2)]^{1/2}$ of the plane
wave near the soliton center over a range $k_s^{-1}$ and a phase
shift $\delta_k(x)=\tan^{-1}((k_s\tanh k_sx)/k)+\tan^{-1}(k_s/k)$.
For $x\rightarrow-\infty$, $s_k(x)\rightarrow 1$; for
$x\rightarrow\infty$,
$s_k(x)\rightarrow(k+ik_s)/(k-ik_s)=\exp(i\delta_k)$, and $s_k(x)$
becomes the usual s-matrix. We note that the bound state solution
and its zero-frequency eigenvalue is contained in the scattering
solution as a pole in the s-matrix for $k\rightarrow ik_s$.

Inserting the gauge transformation in Eq. (\ref{gt}) we obtain for
the zero-mode $\delta{u}\propto\cosh^{-2}(k_sx)\propto\nabla u_s$
which thus corresponds to the translation or Goldstone mode
associated with the position of the soliton. For the extended
states the gauge transformation $U$ provides an envelope of range
$k^{-1}_s$, i.e., $\delta
u\propto\exp(ikx)s_k(x)\cosh^{-1}(k_sx)$. The complete solution of
the mode equation (\ref{lin0}) thus takes the form
\begin{eqnarray}
\delta u = \frac{A}{\cosh^2k_sx} +
\sum_kB_ke^{-\Omega_kt}\frac{e^{ikx}}{\cosh
k_sx}s_k(x),\label{fluc2}
\end{eqnarray}
expressing the fluctuations of the slope field about the static
right hand soliton $u_s$; $A$ and $B_k=B^\ast_{-k}$ are expansion
coefficients. The first term in Eq. (\ref{fluc2}) is the time
independent translation mode. The second term corresponds to a
band of damped localized states with a gap $\nu k^2_s$ in the
spectrum $\Omega_k = \nu(k^2+k^2_s)$. Moreover, the scattering
modes are transparent and phase-shifted by
$\delta_k=2\tan^{-1}(k_s/k)$, implying according to Levinson's
theorem that the band is depleted by precisely one state
corresponding to the translation mode.

In order to make contact with the asymptotic analysis yielding the
spectrum in Eq. (\ref{spec}) we observe that the fluctuations
given by Eq. (\ref{fluc2}) do not exhaust the spectrum. Since the
gauge factor $U=\cosh^{-1}k_sx$ provides a fall-off envelope we
can extend the set of solutions by an analytical continuation in
the wavenumber $k$ in the same manner as the translation mode is
retrieved by setting $k=ik_s$. Thus shifting $k\rightarrow
k+i\kappa$, where $|\kappa|\leq k_s$, we obtain by insertion the
complex spectrum
\begin{eqnarray}
E_{k,\kappa}=\nu(k^2+k_s^2-\kappa^2)+2i\nu k\kappa,~~|\kappa|\leq
k_s, \label{ev2}
\end{eqnarray}
and associated fluctuation modes
\begin{eqnarray}
\delta u = \frac{A}{\cosh^2k_sx}+ \sum_{k,\kappa}B_{k,\kappa}
e^{-(\Omega_k-\nu\kappa^2)t}e^{ik(x-2\nu\kappa t)}
\frac{e^{-\kappa x}}{\cosh k_sx}s_{k+i\kappa}(x),\label{fluc3}
\end{eqnarray}
where $B_{k,\kappa}=B^\ast_{-k,\kappa}$ since $\delta u$ is real.

The expressions (\ref{ev2}) and (\ref{fluc3}) provide the complete
analytically continued solution of the fluctuation spectrum about
the static soliton compatible with the imposed physical boundary
conditions. For $\kappa = \pm k_s$ we recover the spectrum
(\ref{spec}) of right hand and left hand extended gapless modes
propagating towards the soliton center with velocity $v=2\nu
k_s=\lambda u$. For $|\kappa|<k_s$ we obtain a band of gapful
modes with  localized envelopes propagating with velocity
$2\nu\kappa$ towards the soliton center. Finally, for $\kappa =0$
the envelope is symmetric, the spectrum $E_{k,0}$ is real, and the
mode has no propagating component; for $k=0$ and $\kappa=k_s$ we
retrieve the time independent translation mode. With the exception
of the extended mode for $\kappa=\pm k_s$, the envelope modes for
$|\kappa|<k_s$ are {\em dynamically pinned} to the soliton. This
phenomenon of {\em localization} or {\em dynamical pinning} of the
modes is associated with the complex spectrum in Eq. (\ref{ev2})
resulting from the non-Hermitian character of the eigenvalue
problem. In all cases the modes are damped with a damping constant
given by $\nu(k^2+k_s^2-\kappa^2)$. We mention that a
non-Hermitian eigenvalue spectrum is also encountered in the
context of flux pinning and the transverse Meissner effect in high
$T_c$ superconductors \cite{Hatano96,Hatano97}. Here the imaginary
gauge field is uniform and is given by the applied transverse
magnetic field; in the present case the gauge field is spatially
varying and given by the nonlinear soliton excitations.

\begin{figure}
\includegraphics[width=0.6\hsize]
{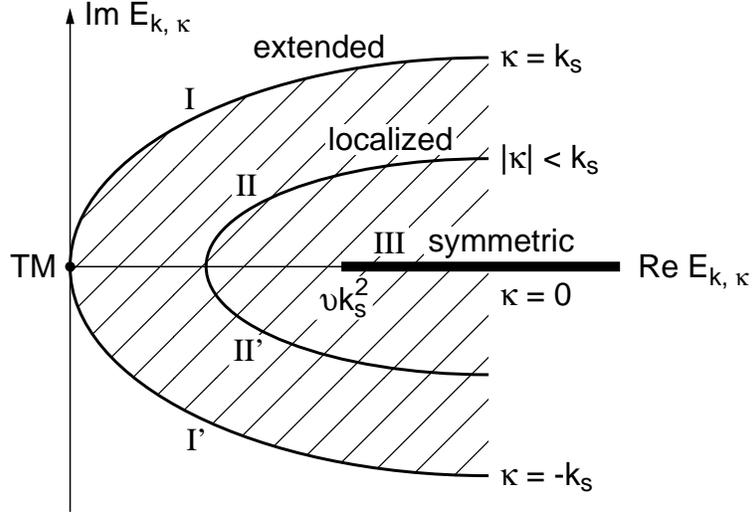} \caption{The complex eigenvalue spectrum for the
fluctuating linear modes,
$\text{Re}E_{k,\kappa}=\nu(k^2+k_s^2-\kappa^2)$ and
$\text{Im}E_{k,\kappa}=2\nu k\kappa$. The bounding parabola for
$\kappa=k_s$ and $\kappa=-k_s$ corresponds to the left and right
extended modes propagating towards the soliton center; they are
denoted I and I', respectively. The shaded area bounded by the
parabola corresponds to localized propagating modes for
$\kappa\neq 0$. For $\kappa=0$ the spectrum is real corresponding
to a localized non-propagating symmetric mode. The point TM
corresponds to the translation mode.} \label{fig7}
\end{figure}

\begin{figure}
\includegraphics[width=0.6\hsize]
{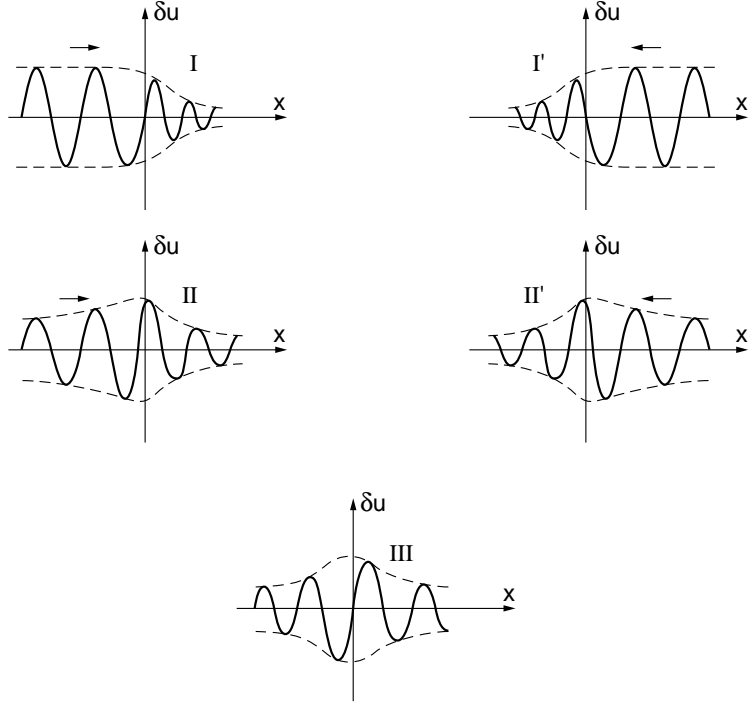} \caption{ The fluctuation patterns of the pinned
dynamical modes corresponding to the sectors of the eigenvalue
spectrum in Fig.~\ref{fig7}. The arrows indicate the propagation
directions.} \label{fig8}
\end{figure}

\begin{figure}
\includegraphics[width=0.6\hsize]
{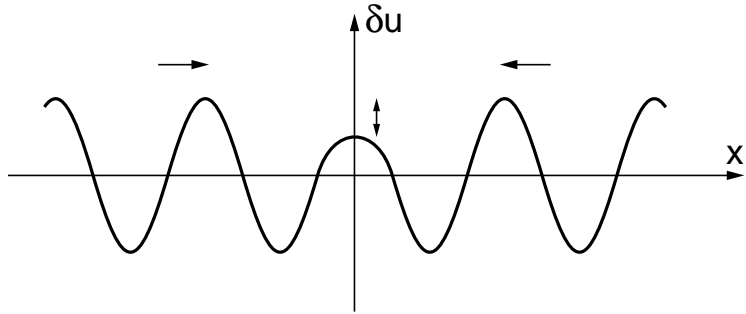} \vspace{1.5cm} \caption{We depict the extended mode
propagating from right and left towards the soliton center which
acts like a sink. The center point at $x=0$ oscillates with
frequency $\omega=kv$. The arrows indicate the propagation
direction.} \label{fig9}
\end{figure}

In Fig.~\ref{fig7} we have depicted the spectrum in a plot of the
imaginary part of $E_{k,\kappa}$ versus its real part. In
Fig.~\ref{fig8} we have shown the associated characteristic
fluctuation mode patterns. Specifically, in order to obtain a real
extended mode propagating towards the soliton center with velocity
$v=2\nu k_s\lambda u$ from the left and with velocity $-v$ from
the right with continuous derivative at $x=0$, thus extending the
asymptotic solution (\ref{fluc1}) to the whole axis we form an
appropriate linear combination from Eq. (\ref{fluc3}) with equal
weights according to the assignments $\pm k$ and $\kappa=\pm k_s$.
Ignoring here the modulation factor $s_{k+i\kappa}(x)$, which is
easily incorporated, we obtain
\begin{eqnarray}
\delta u \propto e^{-\nu k^2t} \frac {e^{-k_sx}\cos
k(x-vt)+e^{k_sx}\cos k(x+vt)}{\cos k_sx}. \label{realfl}
\end{eqnarray}
This mode is depicted in Fig.~\ref{fig9}
%

\subsection{The noisy Burgers case}
In order to discuss the fluctuation spectrum in the noisy case we
must address the coupled field equations (\ref{mf1}) and
(\ref{mf2}) and expand the slope field $u$ and noise field $p$
about the soliton configurations, in the single soliton case given
by Eqs. (\ref{usol}) and (\ref{psol}) and in the multi-soliton
case by Eqs. (\ref{solconf}) and (\ref{noiseconf}). As also
resulting from the analysis in paper II it is convenient to make
use of a symmetrical formulation and introduce the auxiliary noise
field $\varphi$ by means of the shift
\begin{eqnarray}
p=\nu(u-\varphi). \label{snf}
\end{eqnarray}
The field equations then assume the symmetrical form
\begin{eqnarray}
&&\left(\frac{\partial}{\partial t}-\lambda u\nabla\right)u=
\nu\nabla^2\varphi, \label{smf1}
\\
&&\left(\frac{\partial}{\partial t}-\lambda u\nabla\right)\varphi=
\nu\nabla^2u. \label{smf2}
\end{eqnarray}
The single soliton solution $u_s^\mu$ is given by Eq. (\ref{usol})
and the associated noise solution by
\begin{eqnarray}
\varphi_s^\mu=\mu u_s + \text{const}. \label{psisol}
\end{eqnarray}
Expanding about a general multi-soliton configuration
$(u_s,\varphi_s)$, where $u_s$ is given by Eq. (\ref{solconf}) and
$\varphi_s$ by (modulus a constant)
\begin{eqnarray}
\varphi_s(x,t) = \frac{2\nu}{\lambda}\sum_{p=1}^n |k_p|\tanh
|k_p|(x -v_pt-x_p),\label{phiconf}
\end{eqnarray}
by setting $u=u_s+\delta u$ and $\varphi=\varphi_s+\delta\varphi$,
we obtain the coupled linear equations of motion
\begin{eqnarray}
&&\left(\frac{\partial}{\partial t}-\lambda u_s\nabla\right)\delta
u= \nu\nabla^2\delta\varphi + \lambda(\nabla u_s)\delta u,
\label{linu}
\\
&&\left(\frac{\partial}{\partial t}-\lambda
u_s\nabla\right)\delta\varphi= \nu\nabla^2\delta u +
\lambda(\nabla\varphi_s)\delta u, \label{linphi}
\end{eqnarray}
determining the fluctuation spectrum of superimposed linear modes.

The analysis proceeds as in the noiseless case. Referring to Eqs.
(\ref{solconf}) and (\ref{phiconf}) we note that in the
inter-soliton matching regions of constant slope and noise fields
$\nabla u_s=\nabla\varphi_s = 0$. The Eqs. (\ref{linu}) and
(\ref{linphi}) then decouple as in the linear Edwards-Wilkinson
case and setting $u_s=u$ and looking for solutions of the plane
wave form $\delta u,\delta\varphi\propto\exp(-E_kt)\exp(ikx)$ we
obtain $\delta u\pm\delta\varphi\propto \exp(-E^\pm_kt)\exp(ikx)$,
i.e., $\delta u\propto [A\exp(-E_k^+t)+B\exp(-E_k^-t)]\exp(ikx)$,
where the non-Hermitian complex eigenvalue spectrum similar to the
noiseless case is given by
\begin{eqnarray}
E^\pm_k = \pm\nu k^2 - ivk, ~~v=\lambda u. \label{spec2}
\end{eqnarray}
The $\delta u$ mode (and likewise the $\delta\varphi$ mode) thus
corresponds to a propagating wave with both a growing and decaying
component,
\begin{eqnarray}
\delta u\propto(Ae^{-\nu k^2t} + Be^{\nu k^2t})e^{ik(x+vt)}.
\label{fluc5}
\end{eqnarray}
These aspects are consistent with the general phase space behavior
depicted in Fig.~\ref{fig2}, whereas the propagating aspect as in
the noiseless case is the manifestation of a mode transmutation
from diffusive modes in the Edwards-Wilkinson case to propagating
modes in the Burgers case. As indicated in Fig.~\ref{fig4} the
linear mode propagates to the left for $u>0$ and to the right for
$u<0$; for $u=0$ the propagation velocity vanishes and we retrieve
the diffusive modes in the Edwards-Wilkinson case. We note in
particular that for a static right hand soliton with boundary
values $u_\pm=\pm u$, equivalent to the noiseless case discussed
above, the mode propagates towards the soliton center which acts
like a sink; for a static noise-induced left hand soliton with
boundary values $u_\pm=\mp u$ the situation is reversed and the
modes propagate away from the soliton which in this case plays the
role of a source.

In the soliton regions the slope and noise fields vary over a
scale $k_s^{-1}$ and we must address the equations (\ref{linu})
and (\ref{linphi}). Introducing the auxiliary variables
\begin{eqnarray}
\delta X^\pm = \delta u \pm \delta\varphi,\label{auxvar}
\end{eqnarray}
and the general gauge function $g_s$ defined by the slope profile
\begin{eqnarray}
g_s(x,t)=\frac{\lambda}{2\nu}u_s(x,t),\label{gf2}
\end{eqnarray}
the equations (\ref{linu}) and (\ref{linphi}) take the form
\begin{eqnarray}
-\frac{\partial\delta X^\pm}{\partial t} = \pm D(\pm g_s)\delta
X^\pm - \frac{\lambda}{2}(\nabla u_s\pm\nabla\varphi_s)\delta
X^\mp,~~~~ \label{symlin}
\end{eqnarray}
where $D(\pm g_s)$ is the ``gauged'' Schr\"{o}dinger operator
\begin{eqnarray}
D(\pm g_s) = -\nu(\nabla\pm g_s)^2+
\frac{\lambda^2}{4\nu}u^2_s-\frac{\lambda}{2}\nabla\varphi_s,
\label{sham2}
\end{eqnarray}
describing the motion of a particle in the soliton-defined
potential $(\lambda^2/4\nu)u_s^2-(\lambda/2)\nabla\varphi_s$
subject to the gauge field $g_s$. In the regions of constant slope
and noise fields we have $\nabla u_s=\nabla\varphi_s=0$, $u_s=u$,
$g_s=\lambda u/2\nu$, $D(\pm g_s)\rightarrow -\nu(\nabla\pm\lambda
u/2\nu)^2+(\lambda^2/4\nu)u^2$, and searching for solutions of the
form $\delta X^\pm\propto\exp(-E_kt)\exp(ikx)$ we recover the
spectrum in Eq. (\ref{spec2}) and since $\delta u=\delta
X^++\delta X^-$ the mode in Eq. (\ref{fluc5}). In the soliton
regions we have $\nabla\varphi_s^\mu = \mu\nabla u_s^\mu$, where
$\mu=\pm 1$ for the right and left hand solitons, respectively,
and one of the equations (\ref{sham2}) decouple driving the other
equation parametrically.

The analysis proceeds in a manner analogous to the noiseless case.
Introducing the Cole-Hopf transformation
\begin{eqnarray}
U(x,t)=\exp\left(-\int^x g_s(x',t)~dx'\right), \label{gt3}
\end{eqnarray}
and using the identity $(\nabla + g_s)^2=U\nabla^2U^{-1}$ we
arrive at the coupled Hermitian equations
\begin{eqnarray}
-\frac{\partial\delta X^\pm}{\partial t} = \pm U^{\pm 1}
D(0)U^{\mp 1}\delta X^\pm - \frac{\lambda}{2}(\nabla
u_s\pm\nabla\varphi_s)\delta X^\mp, \label{symlin2}
\end{eqnarray}
which are readily analyzed in terms of the spectrum of $D(0)$
summarized in Eqs (\ref{eve2}) to (\ref{smt}). The exponent or
generator in the gauge transformation in Eq. (\ref{gt3}) samples
the area under the slope profile $u_s$ up to the point $x$. For
$x\rightarrow\infty$, $U\rightarrow\exp(-\lambda M/2\nu)$, where
$M$ given by Eq. (\ref{intslope}) is the conserved total area. In
terms of the height field $h$, $u = \nabla h$, $M = h(+L)-h(-L)$
for a finite system and thus equal to the height offset across a
system of size $L$, i.e., a conserved quantity under growth.
Inserting the soliton profile $u_s$ in Eq. (\ref{solconf}) the
transformation $U$ factorizes in contributions from the individual
local solitons, i.e.,
\begin{eqnarray}
U(x,t) = \prod_{p=1}^n U_p(x,t) ^{\text{sign}(k_p)}~,~ U_p(x,t) =
\cosh^{-1} k_p(x-v_p t -x_p).~~ \label{gt4}
\end{eqnarray}
%
\subsubsection{The single soliton case}
Since the formulation of the linear mode problem in terms of Eqs.
(\ref{linu}) and (\ref{linphi}) deriving from the field equations
(\ref{smf1}) and (\ref{smf2}) and yielding Eqs. (\ref{symlin2}) is
entirely Galilean invariant, we can in analyzing the p-th single
soliton segment of the multi-soliton configuration in Eq.
(\ref{solconf}) without loss of generality boost the soliton to a
rest frame with zero velocity. Thus shifting the slope field of
the p-th soliton by the amount $(u_{p+1}+u_p)$ corresponding to
the propagation velocity $v_p=-(\lambda/2)((u_{p+1}+u_p)$, and
assuming for convenience that $x_p=0$ the soliton profile is given
by $u_s^\mu$ in Eq. (\ref{usol}). Hence we obtain
$\nabla\varphi^\mu_s=\mu\nabla u^\mu_s$ and $D(0)$ given by Eq.
(\ref{sham2}). Noting that $U=U_p^\mu$ and $\nabla u^\mu_s=\mu
uk_sU^2_p$ we find for the fluctuations
\begin{eqnarray}
\delta\tilde X^\pm = U_p^{\mp\mu}\delta X^\pm, ~~ U_p =
\cosh^{-1}k_sx, \label{newvar}
\end{eqnarray}
the Hermitian mode equations
\begin{eqnarray}
-\frac{\partial\delta\tilde X^\pm}{\partial t} = \pm
D(0)\delta\tilde X^\pm - \nu k^2_s(\mu\pm 1)\delta\tilde X^\mp,
\label{symlin3}
\end{eqnarray}
which decouple and are readily analyzed by expanding $\delta\tilde
X^\pm$ on the eigenstates of $D(0)$. We obtain
\begin{eqnarray}
&&\delta X^\mu =\frac{2\nu k^2_s A_0^{(\mu)}t + B_0^{(\mu)}}
{\cosh^2 k_s x}+ (A^{(\mu)}_k e^{-\mu\Omega_kt} + B_k^{(\mu)}
e^{\mu\Omega_kt}) \frac{e^{ikx}s_k(x)}{\cosh k_s x},\label{fluc6}
\\
&&\delta X^{-\mu} =A_0^{(\mu)}+\mu B^{(\mu)}_k\frac{\Omega_k}{\nu
k_s^2}e^ {\mu\Omega_kt}e^{ikx} s_k(x)\cosh k_s x,\label{fluc7}
\end{eqnarray}
describing the fluctuations of the slope and noise fields
$\delta u = (\delta X^++\delta X^-)/2$ and
$\varphi = (\delta X^+-\delta X^-)/2$ or $\delta p=\nu\delta X^-$
about the static right hand ($\mu=+1$) and left hand ($\mu=-1$)
solitons. $A_0^{(\mu)}$, $B_0^{(\mu)}$,  $A^{(\mu)}_k$, and
$B^{(\mu)}_k$ are integration constants fixed by the initial
conditions.

The first terms in Eqs. (\ref{fluc6}) and (\ref{fluc7}) are
associated with the soliton translation modes $\delta
X^\mu_{\text{TM}}\propto\nabla u^\mu_s\propto\cosh^{-2}k_s x$
which propagate with constant momentum
$X^{-\mu}_{\text{TM}}\propto A^{\mu}_0$; we note that the soliton
position and soliton momentum are canonically conjugate variables.
The envelope modulated plane wave terms in Eqs. (\ref{fluc6}) and
(\ref{fluc7}) represent the fluctuations about the soliton. Since
the s-matrix $s_k(x)\rightarrow(k+ik_s)/(k-ik_s)=\exp(i\delta_k)$,
$\delta_k=2\tan^{-1}(k_s/k)$ for $x\rightarrow\infty$ the soliton
induced potentials are transparent and the fluctuations pass
through the soliton only subject to the phase shift $\delta_k$ and
a spatial modulation. We also note that Levinson's theorem implies
that the band is depleted by one mode corresponding to the
translation mode \cite{Fogedby85}.

Confining the fluctuations to the noiseless transient submanifold
$p=0$ we have $\delta p=0$, i.e., $\delta u=\delta\varphi$, and we
obtain for the right hand soliton ($\mu=+1$) $\delta X^-=0$ implying
$A_0^{(1)} =0$ and $B_k^{(1)}=0$ and thus
\begin{eqnarray}
\delta X^+=2\delta u = \frac{B_0^{(1)}}{\cosh^2k_sx}
+\frac{A_k^{(1)}e^{-\Omega_k t}e^{ikx}s_k(x)}{\cosh k_sx},
\end{eqnarray}
in accordance with Eq. (\ref{fluc2}) in the noiseless Burgers case,
i.e., a translation mode and a band of damped localized pinned modes.

Likewise, on the noisy stationary submanifold $p=2\nu u$ we require
$\delta p=2\nu\delta u$, i.e., $\delta u=\delta\varphi$, and we
obtain for the noise induced left hand soliton ($\mu=-1$)
$\delta X^+=0$ entailing $A^{(-)} = 0$ and $B_k^{(-)}=0$ and we
have
\begin{eqnarray}
\delta X^-=2\delta u = \frac{B_0^{(-1)}}{\cosh^2k_sx}
+\frac{A_k^{(-1)}e^{\Omega_k t}e^{ikx}s_k(x)}{\cosh k_sx},
\end{eqnarray}
composed of a translation mode and localized modes. However,
unlike the fluctuations about the noiseless right hand soliton
which are damped, the fluctuations associated with the noisy left
hand soliton are growing in time. This behavior is consistent with
the phase space plot in Fig.~\ref{fig2} and the Edwards-Wilkinson
case discussed in Sec.~\ref{growint}.

In general there are also fluctuations perpendicular to the
submanifolds and we are led to consider the coupled equations
(\ref{fluc6}) and (\ref{fluc7}). The fluctuations are modulated by
the gauge factors $\cosh k_s x$ and $\cosh^{-1} k_s x$. As in the
noiseless case the spatial modulation of the plane wave form
allows us to extend the spectrum by an analytical continuation in
the wavenumber $k$ and in this manner match the spectrum to the
inter-soliton regions. In fact, noting that $\delta X^\pm$
according to (\ref{symlin2}) decouples for $\nabla u_s$,
$\nabla\varphi_s\rightarrow 0$ in the inter-soliton regions, we
obtain by setting  $k\rightarrow k\pm ik_s$ the shift
$\Omega_k=\nu(k^2+_s^2)\rightarrow\nu(k^2\pm2ikk_s)$ and
$\exp(ikx)\cosh^\pm k_sx\rightarrow\text{const.}$ and we achieve a
matching to the extended propagating modes. The gauge
transformation in Eq. (\ref{gt4}) thus permits  a complete
analysis of the linear fluctuation spectrum about a multi-soliton
configuration.
\subsubsection{The two-soliton case}
In order to illustrate how the fluctuation spectrum is established
across the soliton configuration and how the matching is
implemented we consider the fluctuations about the two-soliton
configuration in Eq. (\ref{2u}) with associated noise field
\begin{eqnarray}
\varphi_2(x,t) = \frac{u}{2}\left[\tanh\frac{k_s}{2}(x-vt-x_1) +
\tanh\frac{k_s}{2}(x-vt-x_2)\right]. \label{2phi}
\end{eqnarray}
The Hermitian linear mode equations are given by (\ref{symlin2}) with
$u_s=u_2$, $\varphi_s=\varphi_2$ and the  Cole-Hopf transformation
\begin{eqnarray}
U_2(x,t) = \exp\left[-\frac{\lambda}{2\nu}\int^x
u_2(x',t)~dx'\right]. \label{gt5}
\end{eqnarray}
The soliton pair propagates with velocity $v = -\lambda u/2$ so in
order to render the gauge transformation time independent and thus
facilitate the analysis we boost the configuration to a rest
frame, $u_2\rightarrow u_2-u/2$. Inserting $u_2$ we then obtain
the specific gauge transformation
\begin{eqnarray}
U_2(x) = \left[\frac{\cosh\frac{k_s}{2}(x-x_2)}
{\cosh\frac{k_s}{2}(x-x_1)}\right]e^{k_s x/2}, \label{gt6}
\end{eqnarray}
a special case of (\ref{gt4}).

In order to simplify the discussion of the mode equations
(\ref{symlin2}) with $D(0)$ given by (\ref{sham2}) we introduce
the notation $u_\pm=(u/2)\tanh k_s/2(x-x_\pm)$, $x_+=x_1$,
$x_-=x_2$, for the individual solitons contributing to $u_2$ and
$\varphi_2$, i.e., $u_2 =u^1-u^2-u/2$ and $\varphi_2=u^1+u^2$. A
simple estimate of the potential
$(\lambda^2/4\nu)u_2^2-(\lambda/2)\nabla\varphi_2$ in
(\ref{sham2}) then yields $\nu(k_s/2)^2[1-V_+-V_-]$, where we have
also introduced the notation $V_\pm = 2/\cosh^2(k_s/2)(x-x_\pm)$,
for the soliton-induced potentials. Moreover, $(\lambda/2)(\nabla
u_2\pm\nabla\varphi_2)=\pm\nu(k_s/2)V_\pm$, and we arrive at the
two-soliton mode equations
\begin{eqnarray}
-\frac{\partial\delta X^\pm}{\partial t} = \pm U_2^{\pm 1}
D(0)U_2^{\mp 1} \delta X^\pm \mp\nu(k_s/2)^2V_\pm\delta X^\mp,
\label{m}
\end{eqnarray}
with $D(0)$ given by
\begin{eqnarray}
D(0) = -\nu\nabla^2 + \nu(k_s/2)^2(1-V_+-V_-). \label{sham3}
\end{eqnarray}
In the regions of constant slope field $V_\pm\sim 0$ and $\delta
X^\pm$ decouple. To the right of the soliton pair for $x\gg
x_1,x_2$ we have $U_2\propto\exp(k_s(x_1-x_2)/2)\exp(k_sx/2)$ and
we obtain the envelope solutions $\delta
X^\pm\propto\exp[\pm\nu(k^2+(k_s/2)^2)t]U_2^{\pm 1}\exp(ikx)$
which are matched to the asymptotic plane wave solution by setting
$k\rightarrow k\pm ik_s/2$, yielding $\delta
X^\pm\propto\exp(\mp\nu k^2t)\exp(ik(x-\nu k_st))$, i.e., a mode
propagating to the right with velocity $\nu k_s=\lambda u/2=v$. To
the left for $x\ll x_1, x_2$, we have
$U_2\propto\exp(k_s(x_2-x_1)/2)\exp(k_sx/2)$ and correspondingly
the envelope solutions $\delta
X^\pm\propto\exp[\pm(\nu(k^2+(k_s/2)^2)t]U_2^{\pm 1}\exp(ikx)$.
Matching these solutions to the plane wave solutions by the
analytic continuation $k\rightarrow k\pm ik_s/2$, we obtain the
same result as above. We note that the change in $U_2$ across the
pair is given by
$\exp(-k_s(x_2-x_1))=\exp(-(\lambda/2\nu)u(x_2-x_1))
\simeq\exp(-(\lambda/2\nu)M_2)$, where $M_2$ is the area enclosed
by the soliton pair. In the region between the solitons for
$x_1\ll x\ll x_2$, $U_2\sim\exp(k_s(x_1+x_2)/2)\exp(-k_s x/2)$ or
for $k\rightarrow k\mp ik_s/2$ the modes $\delta
X^\pm\propto\exp(\mp\nu k^2t)\exp(ik(x+\nu k_st))$, corresponding
to propagation to the left with velocity $-\nu k_s=-\lambda
u/2=-v$. In the soliton region near $x_+$ we have $V_-\sim 0$.
Ignoring the translation mode and phase shift effects  we thus
have from (\ref{m}) $\delta
X^-\propto\exp(\nu(k^2+(k_s)^2)t)U_2^{-1}\exp(ikx)$. By insertion
in Eq. (\ref{m}) we note that $V_1U_2^{-1} = V_2\exp(-k_s x/2)\sim
0$ and that $\delta X^+$ thus is decoupled from $\delta X^-$,
yielding the solution $\delta
X^+\propto\exp(-\nu(k^2+(k_s/2)^2)t)U_2\exp(ikx)$. A similar
analysis applies in the soliton region near $x_2$.

For the plane wave components alone we then obtain, interpolating
to the whole axis and incorporating the s-matrices according to
Eq. (\ref{smt}),
\begin{eqnarray}
&&\delta X^+\sim e^{-\nu(k^2+(k_s/2)^2)t}
\frac{\cosh(k_s/2)(x-x_2)}{\cosh(k_s/2)(x-x_1)}e^{k_s
x/2}e^{ikx}s_k(x-x_1)s_k(x-x_2),\label{f1}
\\
&&\delta X^-\sim e^{\nu(k^2+(k_s/2)^2)t}
\frac{\cosh(k_s/2)(x-x_1)}{\cosh(k_s/2)(x-x_2)}e^{-k_s
x/2}e^{ikx}s_k(x-x_1)s_k(x-x_2).\label{f2}
\end{eqnarray}
We thus we pick up the phase shift $\delta_k=2\tan^{-1}(k_s/k)$
across each soliton, the plane wave components are moreover
spatially modulated by the gauge transformation
$U_2=\exp[-(\lambda/2\nu)\int u_2dx]$ sampling the area under
$u_2$.

Finally, we boost the mode to the velocity $v=-\nu k_s=-\lambda
u/2\nu$, shift the wavenumber $k\rightarrow k\pm ik_s/2$ for
$\delta X^\pm$ and obtain the modulated plane wave associated with
the propagating two-soliton mode with vanishing boundary
conditions
\begin{eqnarray}
&&\delta X^+\sim e^{-\nu k^2t}
\frac{\cosh(k_s/2)(x-vt-x_2)}{\cosh(k_s/2)(x-vt-x_1)}e^{ikx}
s_{k+ik_s/2}(x-vt-x_1)s_{k+ik_s/2}(x-vt-x_2),~~~~~~~\label{ff1}
\\
&&\delta X^-\sim e^{\nu k^2t}
\frac{\cosh(k_s/2)(x-vt-x_1)}{\cosh(k_s/2)(x-vt-x_2)}e^{ikx}
s_{k-ik_s/2}(x-vt-x_1)s_{k-ik_s/2}(x-vt-x_2).\label{ff2}
\end{eqnarray}
The interpretation of this result is straightforward. In the
regions away form the pair-soliton we obtain plane wave modes with
both a growing and decaying time behavior. Across the propagating
two-soliton configuration the plane wave amplitude and form is
modified by the gauge factor and the s-matrix.
\subsubsection{The multi-soliton case}
In the multi-soliton case the slope configuration $u_s$ and the
associated noise field $\varphi_s$ are given by  Eqs.
(\ref{solconf}) and (\ref{phiconf}). The linear mode problem is
defined by Eqs. (\ref{symlin2}) with the gauge transformation $U$
given by Eq. (\ref{gt3}). As discussed previously the extended
plane wave modes in the inter-soliton regions connecting the
solitons are subject to a mode transmutation to propagating waves
with spectrum given by Eqs. (\ref{spec2}). In the soliton regions
the analysis follows from a generalization of the single and
two-soliton cases discussed above.

A complete analysis is achieved by first noting that $U$ also can
be expressed in the form
\begin{eqnarray}
U(x,t) =
\exp\left[-\frac{\lambda}{2\nu}\int^tdt'~\left(\frac{\lambda}{2}u_s^2(x,t')
+\nu\nabla\varphi_s(x,t')\right)\right], \label{gt7}
\end{eqnarray}
derived by differentiating $U$ in Eq. (\ref{gt3}), using the
equation of motion in Eq. (\ref{smf1}) for the multi-soliton
profile and solving the ensuing differential equation. By
insertion of Eq. (\ref{gt7}) in Eqs. (\ref{symlin2}) we obtain the
linear equations of motion
\begin{eqnarray}
\frac{\partial}{\partial t}\left(U^{\mp 1}\delta X^\pm\right) =
\pm(\nu\nabla^2+\lambda\nabla\varphi_s)U^{\mp 1}\delta X^\pm +
\frac{\lambda}{2}(\nabla u_s\pm\nabla\varphi_s)U^{\mp 1}\delta
X^\mp, \label{finaleq}
\end{eqnarray}
which are readily discussed. In the flat regions $\nabla
u_s=\nabla\varphi_s=0$ , $\delta X^+$ and $\delta X^-$ decouple
and we have
\begin{eqnarray}
\delta X^\pm\propto e^{\mp\nu k^2t} e^{ikx}U^{\pm 1}(x,t),
\label{flat}
\end{eqnarray}
describing a plane wave mode modulated across the soliton regions
by the Cole-Hopf transformation $U(x,t)$ with the explicit form
given by Eq. (\ref{gt4}). Alternatively, we obtain a mode
transmutation to a propagating plane wave between solitons by the
analytical continuation $k\rightarrow k\pm ik_p$ thus absorbing
the spatial modulation in $U(x,t)$ and corroborating the previous
discussion. We note that the form in Eq. (\ref{flat}) is in
accordance with Eqs. (\ref{ff1}) and (\ref{ff2}) in the
two-soliton case. Across the soliton regions $\delta X^+$ and
$\delta X^-$ couple according to Eqs. (\ref{finaleq}). The
analysis in the single soliton case above applies and the plane
wave mode picks up the phase shift $\delta_k$ associated by
Levinson's theorem with the formation of the soliton translation
modes.

%
\section{\label{corr} Statistical properties - Correlations - Scaling}
In this section we discuss the statistical properties of a growing
interface on the basis of the canonical phase space formulation.
Generally, we can express the noisy Burgers equation in Eq.
(\ref{bur}) in the form
\begin{eqnarray}
\frac{\partial u}{\partial t}= \nu\nabla\frac{\delta F}{\delta u}
+ \lambda u\nabla u + \nabla\eta, \label{bur2}
\end{eqnarray}
where the free energy $F$ driving the diffusive term is given by
\begin{eqnarray}
F=\frac{1}{2}\int dx~u^2(x).
\end{eqnarray}
For $\lambda=0$ we have the linear Edwards-Wilkinson equation
describing the temporal fluctuations in a thermodynamic
equilibrium state with temperature $T=\Delta/2\nu k_{\text{B}}$
with stationary distribution given  by Eq. (\ref{stat2}), i.e.,
$P_{\text{st}}\propto\exp[-F/k_{\text{B}}T]$. In the presence of
the nonlinear mode coupling term $\lambda u\nabla u$ Eq.
(\ref{bur2}) does not describe a thermodynamic equilibrium state
but a stationary nonequilibrium state or kinetic growth state. It
is, however, a particular feature of the one dimensional case that
the stationary distribution is known and given by
$P_{\text{st}}\propto\exp[-F/k_{\text{B}}T]$, independent of
$\lambda$. This property also follows from the potential condition
\cite{Deker75,Fogedby80,Stratonovich63}
\begin{eqnarray}
\int dx[\lambda\nabla u-\frac{\lambda\nu}{\Delta}\frac{\delta
F}{\delta u}\nabla(u^2)]=0,
\end{eqnarray}
which is readily satisfied since the integrand becomes a total
derivative. Another way of noting that Eq. (\ref{bur2}) does not
describe an equilibrium state is to express the equation in the
form $\partial u/\partial t=\nu\nabla\delta F'/\delta u +
\nabla\eta$ with effective free energy $F'=(1/2)\int
dx(u^2+(\lambda/3\nu)u^3)$. Apart from the fact that $F'$ includes
an odd power in  $u$ and thus, since $u=\nabla h$, violates parity
invariance it is also unbounded from below for $u\rightarrow
-\infty$ and thus cannot describe a stable thermodynamic state.

The stationary distribution $P_{\text{st}}(u)$ is obtained in the
limit $t\rightarrow\infty$ from the transition probability
$P(u^{\text{i}}\rightarrow u,t)$ for a pathway from the initial
configuration $u^{\text{i}}$ to the final configuration $u$. In
this limit only the linear diffusive modes for $\lambda=0$ persist
characterized by $P_{\text{st}}$. This is consistent with the fact
that the soliton contribution yields $P(u^{\text{i}}\rightarrow
u,t)\propto\exp[-S(t)/\Delta]$, where the action $S(t)$ associated
with the solitonic growth modes, e.g., the two-soliton
configuration, typically grows linearly with $t$ implying that the
contribution to $P$ vanishes.

\subsection{Correlations in the Edwards-Wilkinson case}
In the linear case the correlation function is easily evaluated on the
Langevin level from the stochastic Edwards-Wilkinson equation
(\ref{dif}) and follows directly from Eq. (\ref{ulin}) when averaging
over the noise, see also paper II. In wavenumber-frequency space we
obtain the Lorentzian diffusive form
\begin{eqnarray}
\langle uu\rangle(k,\omega) = \frac{\Delta k^2}{\omega^2+(\nu
k^2)^2}, \label{lcor}
\end{eqnarray}
with diffusive poles at $\omega=\pm i\nu k^2$, strength
$\Delta/(\nu k)^2$ and linewidth $\nu k^2$. We note that both
growing, $u\propto\exp(\nu k^2t)$, and decaying terms,
$u\propto\exp(-\nu k^2t)$, contribute to the stationary
correlations; this is in accordance with the decomposition
(\ref{dec}). In wavenumber-time space we have correspondingly
\begin{eqnarray}
\langle uu\rangle(k,t) = \frac{\Delta}{2\nu}e^{-\nu k^2|t|},
\label{wtc}
\end{eqnarray}
and the diffusive correlations decay on a time scale determined by
$1/\nu k^2$. For the equal-time correlations we obtain in
particular $\langle uu\rangle(k,0)=\Delta/2\nu$, showing the
spatially short ranged correlations in accordance with the
stationary distribution (\ref{stat2}). For later purposes we also
need the spectral form, see paper II,
\begin{eqnarray}
\langle uu\rangle(x,t) = \int\frac{dk}{2\pi}
\frac{\Delta}{2\nu}e^{ikx}e^{-\nu k^2|t|}. \label{specf}
\end{eqnarray}
In order to illustrate the method to be used later in the soliton
case we evaluate here $\langle uu\rangle$ in the linear case on
the basis of the path integral formulation Eq. (\ref{cor}). Since
that the distributions (\ref{stat2}) and (\ref{dis2}) factorize in
wavenumber space we have in a little detail for a system of size
$L$
\begin{eqnarray}
\langle uu\rangle(k,t)\propto
\int\prod_pdu^{\text{i}}_pdu^{\text{f}}_pu^{\text{f}}_ku^{\text{i}}_k
\prod_n\exp\left(-\frac{\omega_n}{\Delta
L}\frac{|u^{\text{f}}_n-u^{\text{i}}_ne^{-\omega_nt}|^2}
{1-e^{-2\omega_nt}}\right) \prod_m\exp¨\left(-\frac{\nu}{\Delta
L}|u^{\text{i}}_m|^2\right).
\end{eqnarray}
Changing variables in $P(u_n^{\text{f}},u_n^{\text{i}},t)$,
$u_n^{\text{f}}-u_n^{\text{i}}\exp(-\omega_nt)\rightarrow u_n$, it
is an easy task to carry out the Gaussian integrals over $u_n$ and
$u_n^{\text{i}}$ and retrieve $\langle uu\rangle(k,t)$ in Eq.
(\ref{wtc}); the evaluation of $\langle uu\rangle$ in the
corresponding harmonic oscillator quantum field calculation was
performed in paper II. Finally, evaluating Eq. (\ref{specf}) we
infer the scaling form (\ref{scal}) with roughness exponent $\zeta
= 1/2$, dynamic exponent $z=2$, and scaling function, see also
paper II,
\begin{eqnarray}
\tilde F(w) = \frac{\Delta}{2\nu}[4\pi\nu]^{-1/2}w^{-1/2}
e^{-1/4\nu w}, w=t/x^z, \label{sfun}
\end{eqnarray}
defining the Edwards-Wilkinson universality class.

Summarizing, the Edwards-Wilkinson  equation describes a
thermodynamic equilibrium state. The dynamical equilibrium
fluctuations are characterized by the gapless dispersion law
$\omega=\nu k^2$. The modes are extensive and diffusive and
controlled by the characteristic decay time $1/\nu k^2$, depending
on the wavenumber $k$.
\subsection{Switching and pathways in the Burgers-KPZ case}
Before we turn to the correlations in the nonlinear Burgers-KPZ
case it is instructive to extract a couple of simple qualitative
consequences of the dynamical approach. As discussed in
Sec.~\ref{growint} the propagation of a two-soliton configuration
constitutes a simple growth situation where at each passage of the
soliton pair the interface grows by a layer. Considering a pair
configuration of amplitude $u$ and size $\ell$ the propagation
velocity is $v=-\lambda u/2$ and the associated action
$S_1=(1/6)\nu\lambda|u|^3T$. Across a system of size $L$ we have
$|v|=L/T$, where $T$ is the switching time, i.e., $|u|=2L/\lambda
T$. For the action associated with the transition pathway of
adding a layer of thickness $h= \int^x u(x',t)dt=|u|\ell=
2L\ell/\lambda T$ we then have
\begin{eqnarray}
S_1(T)=\frac{4\nu}{3\lambda^2}\frac{L^3}{T^2}.
\end{eqnarray}
We note that the thickness $h$ does also depend on the
pair-soliton size $\ell$ which does not enter in the action.
However, the multiplicity or density of soliton pair which enters
in the prefactor of the transition probability must depend on
$1/\ell$ and we obtain qualitatively
\begin{eqnarray}
P\propto \ell^{-1}\exp\left(-\frac{1}{\Delta}\frac{4\nu
L^3}{3\lambda^2T^2}\right).
\end{eqnarray}
In the thermodynamic limit $L\rightarrow\infty$, $P\rightarrow 0$
and the switching via a two-soliton pathway is suppressed. At long
times the action falls off as $1/T^2$.

\begin{figure}
\includegraphics[width=0.6\hsize]
{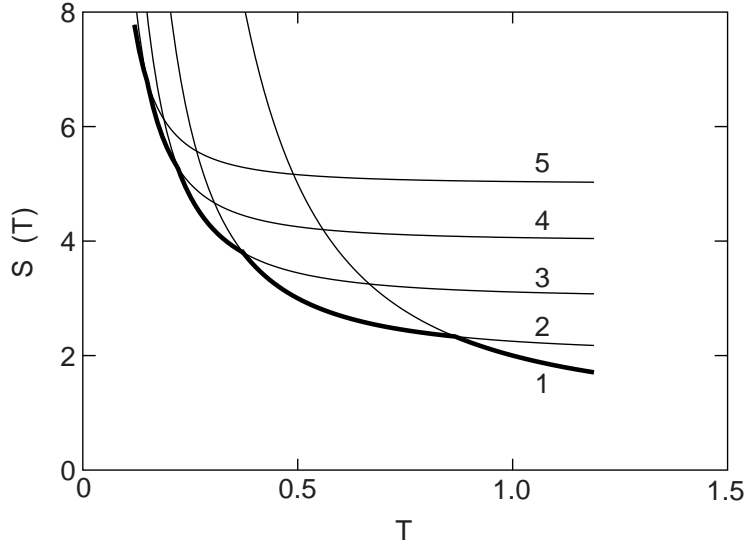} \vspace{1.5cm} \caption{The action given by Eq.
(\ref{switch}) is plotted as a function of $T$ for transition
pathways involving up to $n=5$ soliton pairs. The lowest action
and thus the most probable transition is associated with an
increasing number of soliton pairs at shorter times, indicated by
the heavy limiting curve. The curves are plotted in arbitrary
units.} \label{fig10}
\end{figure}

In the case of a switching pathway by means of two equal amplitude
pair-solitons we obtain, correspondingly, noting that the pairs
propagate with half the velocity, the action $S_2=(1/4)S_1$.
Introducing heuristically a constant nucleation action
$S_{\text{nucl}}$ associated with the noise-induced formation of a
pair, i.e., the appropriate assignment of the noise field $p$, we
have the general expression for the action associated with $n$
pairs
\begin{eqnarray}
S_n(T)=nS_{\text{nucl}}+\frac{1}{n^2}\frac{4\nu}{3\lambda^2}\frac{L^3}{T^2}.
\label{switch}
\end{eqnarray}
In Fig.~\ref{fig10} we have plotted $S$ versus $T$ for $n=1-5$
soliton pairs. Since the curves intersect we infer that the
switching at long times takes place via a single soliton pair. At
shorter times a switching takes place and the transition pathway
proceeds by the excitation of multi-pair-solitons. This is clearly
a finite size effect.

A similar analysis of the soliton switching pathways in the case
of the noise driven Ginzburg-Landau equation has recently been
carried out. Here the analysis, corroborating recent numerical
optimization studies, is simpler because the soliton excitations
are topological and have a fixed amplitude \cite{Fogedby03a}.
\subsection{Anomalous diffusion of growth modes in the Burgers-KPZ case}
On the Langevin level the growth of the interface is a stochastic
phenomena driven by noise. Parameterizing the growth in terms of
growth modes corresponding to the propagation of a gas of
independent pair-solitons in the slope field the dynamical
approach allows a simple interpretation. Noting that the action
associated with the pair mode is given by
$S=(1/6)\nu\lambda|u|^3t$ and denoting the center of mass of the
pair mode by $x=(x_1+x_2)/2$ we have $u=2v/\lambda=2x/t\lambda$
and we obtain  the transition probability
\begin{eqnarray}
P(x,t)\propto\exp\left(-\frac{4}{3}\frac{\nu}{\Delta\lambda^2}
\frac{x^3}{t^2}\right), \label{rwpm}
\end{eqnarray}
for the random walk of independent pair-solitons or steps in the
height profile. Comparing (\ref{rwpm}) with the distribution for
ordinary random walk originating from the Langevin equation
$dx/dt=\eta, \langle\eta\eta\rangle(t)=\Delta\delta(t)$,
$P(x,t)\propto\exp(-x^2/2\Delta t)$, we conclude that the growth
mode performs anomalous diffusion. Assuming pairs of the same
average size, the distribution (\ref{rwpm}) also implies the
soliton mean square displacement,
\begin{eqnarray}
\langle x^2\rangle(t)\propto
\left(\frac{\Delta\lambda^2}{\nu}\right)^{1/z}t^{2/z}~,
\label{msd}
\end{eqnarray}
with dynamic exponent $z=3/2$, identical to the dynamic exponent
defining the KPZ universality class. This result should be
contrasted with the mean square displacement $\langle
x^2\rangle\propto\Delta t^{2/z}$, $z=2$, for ordinary random walk.
The growth modes thus perform superdiffusion. This result is also
obtained using the mapping of the KPZ equation to directed
polymers in a random medium \cite{Halpin95}.

The diffusion of solitons or growth modes is another signature of
the stationary nonequilibrium state. Whereas the extended
diffusive equilibrium modes for a particular wavenumber $k$ are
characterized by the stationary distribution
$P_{\text{st}}\propto\exp(-(\nu/\Delta L)|u_k|^2)$, the random
walk distribution of the growth modes $P(x,t)\propto
t^{-2/3}\exp({\text{const}\times x^3/t^2})$ vanishes for
$t\rightarrow\infty$. The growth modes  or solitons disperse
diffusively over the system and generate the stationary growth.
\subsection{Correlations in the Burgers-KPZ case - general}
As regard the correlations in the nonlinear Burgers-KPZ case the
situation is more complex. The noisy Burgers equation (\ref{bur})
is not easily amenable to a direct analysis of the noise averaged
correlations and we limit ourselves to a discussion of $\langle
uu\rangle(x,t)$ within the canonical phase space approach. In
order to evaluate the slope correlations $\langle uu\rangle(x,t)$
by means of Eq. (\ref{cor}), i.e.,
\begin{eqnarray}
\langle uu\rangle(x,t) = \int\Pi
du^{\text{i}}du^{\text{f}}~u^{\text{f}}(x)u^{\text{i}}(0)
P(u^{\text{f}}, u^{\text{i}},t)P_{\text{st}}(u^{\text{i}}),
\label{cor1}
\end{eqnarray}
we note that the basic ingredient is the transition probability
$P(u^{\text{f}},u^{\text{i}},t)$ from an initial configuration
$u^{\text{i}}$ at time $t=0$ to a final configuration
$u^{\text{f}}$ at time $t$.
\subsubsection{Sum rule}
Before continuing we observe that in the short time limit
$t\rightarrow 0$ it follows from the definition that
$P(u^{\text{f}},u^{\text{i}},t)\rightarrow
\delta(u^{\text{f}}-u^{\text{i}})$. The equal time correlations
are thus determined by the stationary distribution
$P_{\text{st}}(u^{\text{i}})$ given by Eq. (\ref{stat2}),
\begin{eqnarray}
P_{\text{st}}(u^{\text{i}})\propto\exp\left[-\frac{\nu}{\Delta}\int
dx~u^{\text{i}}(x)^2\right], \label{stat3}
\end{eqnarray}
and we have in wavenumber space $\langle
uu\rangle(k,0)=\Delta/2\nu$. In wavenumber-frequency space we thus
infer the general sum rule
\begin{eqnarray}
\int\frac{d\omega}{2\pi}\langle uu\rangle(k,\omega) =
\frac{\Delta}{2\nu}. \label{sumrule}
\end{eqnarray}
The sum rule is independent of the presence of the nonlinear
growth term $\lambda u\nabla u$ and thus is another consequence of
the static fluctuation dissipation theorem which holds for the
Burgers-KPZ equations \cite{Huse85,Halpin95}.
\subsubsection{The transition probability}
As discussed in Sec.~\ref{growint} the working hypothesis is that
a growing interface at a particular time instant can be
represented by a dilute gas of matched localized soliton
excitations or growth modes with superimposed linear extended
diffusive modes. From the analysis in Sec.~\ref{fluc} we thus have
\begin{eqnarray}
&&u(x,t)=u_s(x,t)+\delta u(x,t), \label{uu}
\\
&&p(x,t)=p_s(x,t)+\delta p(x,t), \label{pp}
\end{eqnarray}
where $u_s$ and $p_s$ (or $\varphi_s$) are given by the
multi-soliton representations in  Eqs. (\ref{solconf}) and
(\ref{noiseconf}) (or for $\varphi_s$ in Eq. (\ref{phiconf})). In
the flat regions for constant slope $\delta u = (1/2)(\delta
X^++\delta X^-)$ and $\delta p=\nu\delta X^-$ are given by Eqs.
(\ref{flat}) (across the soliton regions $\delta u$ and $\delta p$
vary in a more complicated manner as discussed in
Sec.~\ref{fluc}).

Inserting Eqs. (\ref{uu}) and (\ref{pp}) in Eq. (\ref{act}) and
using the equation of motion (\ref{mf1}) the action $S$ decomposes
in a soliton contribution $S_{\text{sol}}$ and a linear
contribution $S_{\text{lin}}$, $S=S_{\text{sol}}+S_{\text{lin}}$,
where
\begin{eqnarray}
&&S_{\text{sol}}= \frac{1}{2}\int dxdt(\nabla p_s)^2,
\\
&&S_{\text{lin}}= \frac{1}{2}\int dxdt(\nabla\delta p)^2.
\end{eqnarray}
This decomposition implies that the transition probability
$P(u^{\text{f}},u^{\text{i}},t)$ accordingly factorizes like
\begin{eqnarray}
P(u^{\text{f}},u^{\text{i}},t)=
P_{\text{sol}}(u_s^{\text{f}},u_s^{\text{i}},t)
P_{\text{lin}}(\delta u^{\text{f}},\delta u^{\text{i}},t),
\label{fact}
\end{eqnarray}
where $P_{\text{sol}}\propto\exp(-S_{\text{sol}}/\Delta)$ and
$P_{\text{lin}}\propto\exp(-S_{\text{lin}}/\Delta)$. Disregarding
phase shift effects and amplitude modulations due to the dilute
soliton gas, $P_{\text{lin}}$ can be worked out as in the
Edwards-Wilkinson case in Sec. \ref{growint}, yielding the
expression (in wavenumber space)
\begin{eqnarray}
P(\delta u^{\text{i}},\delta u^{\text{f}},t)\propto
\exp\left[-\frac{\nu}{\Delta}\int\frac{dk}{2\pi} \frac{ |\delta
u^{\text{f}}_k-\delta
u^{\text{i}}_k\exp(-\omega_kt)|^2}{1-\exp(-2\omega_kt)}\right],
\end{eqnarray}
with limiting distribution $P_{\text{st}}(\delta
u^{\text{f}})\propto\exp[-(\nu/\Delta)\int(dk/2\pi)|\delta
u_k^{\text{f}}|^2]$ for $t\rightarrow\infty$.

For the soliton part we obtain inserting Eq. (\ref{actsol})
\begin{eqnarray}
P(u^{\text{i}},u^{\text{f}},t)\propto \exp\left[-\frac{\nu\lambda
t}{6\Delta}\sum_{p=1}^n|u_{p+1} - u_p|^3\theta(u_p-u_{p+1})
\right], \label{dissol}
\end{eqnarray}
in terms of the soliton boundary values $u_p$ as depicted in
Fig.~\ref{fig4}. Note also that only the noise induced left hand
solitons contribute to the action. We stress that the expression
(\ref{dissol}) by construction only holds in-between soliton
collisions. In fact, at long times the appropriate expression for
$P(u^{\text{i}},u^{\text{f}},t)$ must approach
$P_{\text{st}}\propto\exp[-(\nu/\Delta)\int dx
u^{\text{f}}_s(x)^2]$ in accordance with Eq. (\ref{fact}).
Likewise, the correct expression for the multi-soliton energy must
vanish in the long time limit corresponding to the migration of
the phase space orbit to the transient and stationary zero energy
submanifolds $p=0$ and $p=2\nu u$, as discussed in paper III.
\subsubsection{Multi-soliton correlations - Scaling properties}
Inserting Eq. (\ref{uu}) in Eq. (\ref{cor1}) the slope
correlations separate in a soliton part and a linear (diffusive)
part,
\begin{eqnarray}
\langle uu\rangle(x,t)=\langle u_su_s\rangle(x,t)+\langle\delta
u\delta u\rangle(x,t).
\end{eqnarray}
Apart from phase shift and amplitude modulation effetcs due to the
dilute soliton gas, the linear or diffusive correlations $\langle
\delta u\delta u\rangle$ basically have the form given by Eq.
(\ref{specf}). For the soliton contribution $\langle
u_su_s\rangle$ we obtain, inserting $u_s$ from Eq.
(\ref{solconf}), $P(u^{\text{f}}_s,u^{\text{i}},t)$ from Eq.
(\ref{dissol}), and for the stationary distribution
\begin{eqnarray}
P_{\text{st}}\propto\exp\left[-\frac{\nu}{\Delta}\sum_{p=1}^n
u_p^2|x_p-x_{p-1}|\right]=
\exp\left[\frac{8\nu^3}{\Delta\lambda^2}\sum_{p\neq
p'}k_pk_{p'}|x_p-x_{p'}|\right], \label{solstatdis}
\end{eqnarray}
and moreover introducing the soliton amplitude $k_p =
(\lambda/4\nu)(u_{p+1}-u_p)$ from Eq. (\ref{mampl}),
\begin{eqnarray}
\langle u_su_s\rangle(x,t)\Omega(t)=&&\sum_{p,q}\int\prod_l
dk_ldv_ldx_l k_pk_q\tanh|k_p|(x_p+v_pt-x)\tanh|k_q|x_q\times
\nonumber
\\
&&\prod_n\exp\left[
-\frac{32\nu^4}{3\Delta\lambda^4}t|k_n|^3\theta(-k_n)\right]
\prod_{n\neq
n'}\exp\left[\frac{8\nu^3}{\Delta\lambda^2}k_nk_{n'}(x_n-x_{n'})\right].~~~~
\label{multicor}
\end{eqnarray}
This formula expresses the contribution to the slope correlations
from a multi-soliton configuration. It follows from the derivation
that the expression only holds for times short compared to the
soliton collision time. The initial configuration $u_s^{\text{i}}$
at time $t=0$ propagates during time $t$ to the the final
configuration $u_s^{\text{f}}$. The associated transition
probability is given by Eq. (\ref{dissol}) and the stationary
distribution by Eq. (\ref{solstatdis}). The integration over
initial and final configurations is effectuated by integrating
over the amplitudes $k_p$, the velocities $v_p$ and the soliton
positions $x_p$ over a system of size $L$. Note that $k_p$
together with $v_p=-(\lambda/2)(u_{p+1}+u_p)$ determine the slope
$u_p$. Likewise the dynamic partition function $\Omega(t)$ is
given by
\begin{eqnarray}
\Omega(t) =\int\prod_l dk_ldv_ldx_l\prod_n\exp\left[
-\frac{32\nu^4}{3\Delta\lambda^4}t|k_n|^3\theta(-k_n)\right]
\prod_{n\neq
n'}\exp\left[\frac{8\nu^3}{\Delta\lambda^2}k_nk_{n'}(x_n-x_{n'})\right].~~
\label{multipar}
\end{eqnarray}
The complex form of Eqs. (\ref{multicor}) and (\ref{multipar})
have so far precluded a more detailed analysis. We can, however,
in the limit of small damping extract the general scaling
properties. For $\nu\rightarrow 0$ we have $k_p\rightarrow\infty$
and the soliton profile given by $\tanh|k_p|(x-v_pt-x_p)$
converges to the sign function $\text{sgn}(x-v_pt-x_p)$,
corresponding to a sharp shock wave. By inspection of Eq.
(\ref{multicor}) we note that a change of the length scale by a
factor $\mu$, i.e., $x\rightarrow\mu x$ and $x_p\rightarrow\mu
x_p$, can be absorbed by a change of the integration variable
$k_p$, $k_p\rightarrow\mu^{-1/2}k_p$. In the action term this
change of $k_p$ is finally absorbed by the scale transformation
$t\rightarrow\mu^{3/2}$. Consequently, for $\nu\rightarrow 0$ we
have $\langle u_su_s\rangle(x,t)=\tilde F(t/x^{3/2})$ in
accordance with the general scaling form in Eq. (\ref{scal}).
\subsection{Correlations in the Burgers-KPZ case - the two-soliton sector}
In the weak noise limit $\Delta\rightarrow 0$ the action in Eq.
(\ref{dis}) provides a general selection criterion determining the
dominant dynamical configuration contributing to the distribution
$P$. In the present section we propose that part of the leading
growth morphology is constituted by a gas of two-soliton or pair
configurations already analyzed in Sec.~\ref{growint}. In  our
numerical studies we have demonstrated that in the limit
$\nu\rightarrow 0$ the pair configuration does constitute a long
lived quasi-particle \cite{Fogedby02a}.

\begin{figure}
\includegraphics[width=0.6\hsize]
{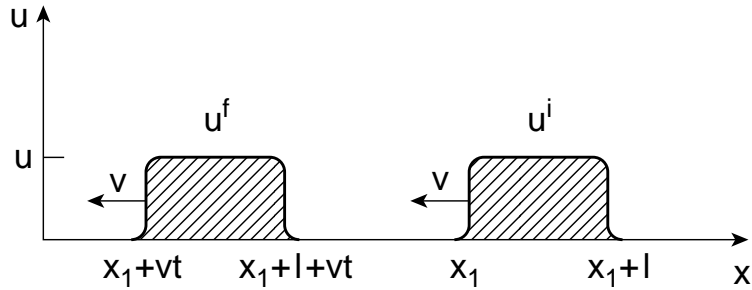} \caption{ We depict the two-soliton configuration in
the limit $\nu\rightarrow 0$ contributing to the slope
correlations $\langle uu\rangle$. The initial pair $u^{\text{i}}$
propagates to the final configuration $u^{\text{f}}$ in time $t$
with velocity $v=-\lambda u/2$.} \label{fig11}
\end{figure}

\begin{figure}
\includegraphics[width=0.6\hsize]
{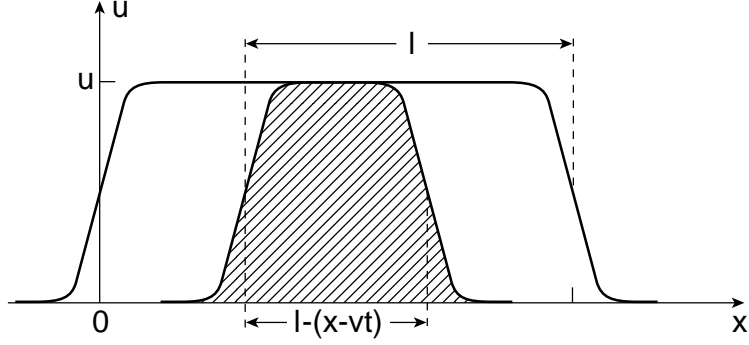} \vspace{1.5cm} \caption{ The two-soliton
configuration of size $\ell = |x_1-x_2|$ and amplitude $u$. The
shaded overlap area of size $2\ell - x$ yields a contribution to
the slope correlation function.} \label{fig12}
\end{figure}

The evaluation of the contribution to the slope correlations from
the two-soliton sector is straightforward and will permit a more
detailed scaling analysis. Specializing the general expression in
Eqs. (\ref{multicor}) and (\ref{multipar}) to the case of two
solitons, i.e., a pair-soliton excitation, noting that
$k_1=-k_2=(\lambda/4\nu)u$, $u_2=u~~ (u_1=u_3=0)$, and
$v_1=v_2=v=-(\lambda/2)u$, and moreover considering the limit of
small $\nu$, or, alternatively, using the expressions pertaining
to the two-soliton case discussed in Sec.~\ref{growint}, we have
\begin{eqnarray}
\langle
uu\rangle(x,t)\Omega_2(t)=&&\left(\frac{\lambda}{4\nu}\right)^2\int
du dx_1dx_2u^2[\text{sign}(x_1)-\text{sign}(x_2)]\times \nonumber
\\
&&[\text{sign}(x_1-x-vt)-\text{sign}(x_2-x-vt)]\times
\nonumber
\\
&&\exp\left[-\frac{\nu\lambda}{6\Delta}t|u|^3\right]
\exp\left[-\frac{\nu}{\Delta}u^2|x_2-x_1|\right],
\end{eqnarray}
with dynamic partition function
\begin{eqnarray}
\Omega_2(t)=\int du dx_1dx_2
\exp\left[-\frac{\nu\lambda}{6\Delta}t|u|^3\right]
\exp\left[-\frac{\nu}{\Delta}u^2|x_2-x_1|\right].
\end{eqnarray}
We note that the final configuration $u^{\text{f}}$ is simply the
initial two-soliton configuration $u^{\text{i}}$ displaced $vt$
along the axis without change of shape. This dynamical evolution
is depicted in Fig.~\ref{fig11}.
The integration over initial and final configurations is carried
out by integrating over the soliton amplitude $u$,
$-\infty<u<\infty$ and the soliton positions $x_1$ and $x_2$ over
a system of size $L$. The integration over the amplitude only
contributes to integral when the pair-solitons overlap, as
indicated in Fig.~\ref{fig12}, and we obtain by inspection of the
overlap contribution, setting $z=x-vt$ and $\ell=|x_2-x_1|$, the
conditions $x_1<z$, $x_1>z-\ell$, $x_1<0$, and $x_1>-\ell$. For
$z>0$, i.e., $x-vt>0$ we have the overlap conditions $0<z<\ell$
and $z-\ell<x_1<0$; for $z<0$ we obtain $-\ell<z<0$ and
$-\ell<x_1<z$. Finally, integrating over the soliton position
$x_1$ and the soliton pair size $\ell$ we arrive at the expression
\begin{eqnarray}
\langle uu\rangle(x,t)=\frac{1}{L} \frac{\int du~u^2
\exp{\left[-\frac{\nu\lambda}{6\Delta}|u|^3t\right]}
\exp{\left[-\frac{\nu}{\Delta}|x-vt|u^2\right]}C^{(1)}_L(u)} {\int
du\exp{\left[-\frac{\nu\lambda}{6\Delta}|u|^3t\right]}
C^{(2)}_L(u)}, \label{corsol2}
\end{eqnarray}
where the cut-off functions
$C^{(1)}_L=\int_0^Ld\ell\ell\exp(-(\nu/\Delta)u^2\ell)$ and
$C^{(2)}_L=\int_0^Ld\ell\exp(-(\nu/\Delta)u^2\ell)$ follow from
the overlap; explicitly they are given by
\begin{eqnarray}
&&C^{(1)}_L(u)= \left(\frac{\Delta}{\nu}\right)^2\frac{1}{u^4}
\left[1-\left(1+\frac{\nu}{\Delta}u^2L\right)
\exp{\left[-\frac{\nu}{\Delta}u^2L\right]} \right],
\\
&&C^{(2)}_L(u)=\left(\frac{\Delta}{\nu}\right)\frac{1}{u^2}
\left[1-\exp{\left[-\frac{\nu}{\Delta}u^2L\right]}\right].
\end{eqnarray}
The overall factor $1/L$ reflects the weight of a single
pair-soliton contribution to the correlations function. In the
thermodynamic limit $L\rightarrow\infty$ this contribution
vanishes. For a dilute gas of pair-solitons of density $n$ we
expect $1/L$ to be replaced by $n$. On the other hand, the further
$L$ dependence of the cut-off functions, is a feature of the
extended nature of the pair-soliton already discussed in
Sec.~\ref{growint}. Both $C^{(1)}$ and $C^{(2)}$ vanish as a
function of $u$  over a scale set by $\sqrt{\Delta/\nu L}$. For
$u\rightarrow\infty$ $C^{(1)}\sim 1/u^4$ and $C^{(2)}\sim 1/u^2$;
for $u=0$ we have $C^{(1)}=L^2/2$ and $C^{(2)}=L$.
\subsection{General scaling properties}
The last issue we deal with is the scaling properties of a growing
interface. The dynamical scaling hypothesis
\cite{Kardar86,Halpin95} and general arguments based on the
renormalization group fixed-point structure \cite{Hwa91,Frey96}
imply the following long time - large distance form of the slope
correlations in the stationary state:
\begin{eqnarray}
\langle uu\rangle(x,t)=(\Delta/2\nu)x^{2\zeta-1}F(x/\xi(t)).
\label{scalform}
\end{eqnarray}
Here $F$ is the scaling function and the roughness exponent $\zeta
$ follows from the explicitly known stationary distribution in Eq.
(\ref{stat2}), the fluctuation dissipation theorem. Within the
canonical phase space approach the stationary distribution follows
from the structure of the zero-energy manifolds which attract the
phase space orbits in the long time limit $t\rightarrow\infty$,
see paper III. The dynamic exponent $z=3/2$ is inferred from the
gapless soliton dispersion law in Eq. (\ref{soldisp}), see paper
II. Since the formulation is entirely Galilean invariant the
exponent $z$ also follows from the scaling law $\zeta+z=2$ in Eq.
(\ref{gal}).

\begin{figure}
\includegraphics[width=0.6\hsize]
{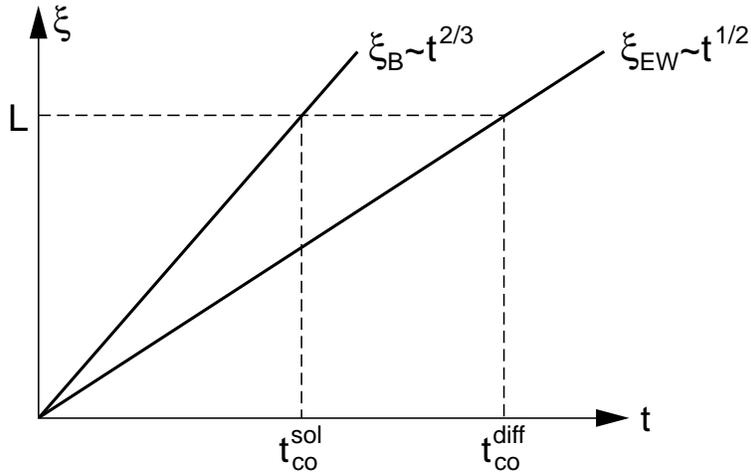} \caption {The correlation lengths
$\xi_{\text{EW}}(t) = (\nu t)^{1/2}$ and $\xi_{\text{B}}(t) =
\left(\Delta/\nu\right)^{1/3}(\lambda t)^{2/3}$ as functions of
$t$. For a finite system of size $L$ the correlation lengths
define the crossover times $t_{\text{co}}^{\text{diff}}\propto
L^2/\nu$ and
$t_{\text{co}}^{\text{sol}}\propto\lambda^{-1}(\nu/\Delta)^{1/2}L^{3/2}$,
determining the transition from transient to stationary growth.}
\label{fig13}
\end{figure}

The lateral growth of fluctuations along the interface  is
conveniently characterized by the time dependent correlation
length $\xi(t)$. Note that for a finite system of size $L$ the
correlation length saturates at the crossover or saturation time
$t_{\text{co}}$ determined by $\xi(t_{\text{co}})=L$. In the
linear Edwards-Wilkinson case $\xi(t)$ characterizes the growth of
diffusive modes and has the form $\xi(t)=(\nu t)^{1/2}$,
consistent with the spectral form in Eq. (\ref{specf}). In the
Burgers-KPZ case $\xi(t)$ describes the propagation of soliton
modes and is given by $\xi(t)=(\Delta/\nu)^{1/3}(\lambda
t)^{2/3}$. The limiting form of the scaling function
$\lim_{w\rightarrow\infty}F(w) = 1$ for $x\gg\xi(t)$. In the
dynamical regime for $\xi(t)\gg x$ the correlation decay, i.e.,
$\langle uu\rangle(x,t) \rightarrow\langle u\rangle \langle
u\rangle = 0$, and the scaling function vanishes like $F(w)\propto
w^{2(1-\zeta)}$ for $w\rightarrow 0$.  In Fig.~\ref{fig13} we have
depicted the correlation length $\xi(t)$ as a function of time for
a system of size $L$, indicating the crossover behavior in the
Edwards-Wilkinson and Burgers-KPZ cases. In Fig.~\ref{fig13} we
have plotted the time scale $T$ as a function of system size
indicating the various dynamic regimes.

%
\subsection{Scaling properties in the two-soliton sector}
In discussing the scaling properties associated with the
two-soliton sector it is convenient to introduce the model
parameters $\ell_0$, setting the microscopic length scale, $t_0$,
setting the microscopic  time scale, $t_{\text{co}}$ defining the
crossover or saturation time for a system of size $L$, and the
correlation length $\xi(t)$,note that $\lambda=\ell_0/t_0$,
\begin{eqnarray}
&&\ell_0=\frac{\Delta}{\nu},
\\
&&t_0=\frac{\Delta}{\nu\lambda},
\\
&&t_{\text{co}}=t_0(L/\ell_0)^{3/2},
\\
&&\xi(t)=\ell_0(t/t_0)^{2/3}.
\end{eqnarray}
Rescaling the amplitude variable $u$ we can then express the pair
correlations in the form
\begin{eqnarray}
\langle uu\rangle(xt) = \frac{\ell_0}{L} \frac { \int du
\exp{\left[-\frac{4}{3}|u|^3\frac{t}{t_{\text{co}}}\right]}
\exp{\left[-4u^2|\frac{x}{L}+u\frac{t}{t_{\text{co}}}|\right]}
F_1(u) } { \int du
\exp{\left[-\frac{4}{3}|u|^3\frac{t}{t_{\text{co}}}\right]} F_2(u)
}, \label{cor2}
\end{eqnarray}
where the cut-off functions originating from the overlap are given
by
\begin{eqnarray}
&&F_1(u)=\frac{1}{4u^2}-\left(1+\frac{1}{4u^2}\right)\exp(-4u^2),
\\
&&F_2(u) = \frac{1}{4u^2}\left(1-\exp(-4u^2)\right).
\end{eqnarray}
The expression (\ref{cor2}) holds for $t>0$ and is even in $x$
(seen by changing $u$ to $-u$). It samples the soliton pair
propagating with velocity $\lambda u/2$ and is in general
agreement with spectral form discussed in the quantum treatment in
paper II.

The weight of a single soliton pair is of order $1/L$ and the
correlation function $\langle uu\rangle$ thus vanishes in the
thermodynamic limit $L\rightarrow\infty$. For a finite system $L$
enters setting a length scale together with the saturation time
$t_{\text{co}}$ defining a time scale, and $\langle uu\rangle$ is
a function of $x/L$ and $t/t_{\text{co}}$ as is the case for the
two-soliton expression (\ref{cor2}). It is instructive to compare
this dependence with the wavenumber decomposition of $\langle
uu\rangle$ in the linear diffusive case for $\lambda=0$. Here
$\langle uu\rangle(xt)\propto (1/L)\sum_{n\neq 0}\exp(-(2\pi
n)^2t/L^2)\exp(i\pi nx/L)$, depending on $x/L$ and $t/L^2$,
corresponding to the saturation time $t_{\text{co}}\propto L^2$,
$z=2$. Keeping only one mode for $n=1$ the correlations $\langle
uu\rangle$ has the same structure as in the soliton case. In the
linear case we can, of course, sum over the totality of modes and
in the thermodynamic limit $L\rightarrow\infty$ replace
$(1/L)\sum_n$ by $\int dk/2\pi$ obtaining the intensive
correlations $\langle uu\rangle(xt)=(\Delta/2\nu)(4\pi\nu
t)^{-1/2}\exp(-x^2/2\nu t)$. Similarly, we expect the inclusion of
multi-soliton modes to allow the thermodynamic limit to be carried
out yielding an intensive correlation function in the Burgers
case.

For a finite system we have in general \cite{Krug92} $\langle
uu\rangle(xt)=(1/L)G_L(x/L,t/L^{3/2}) $ with scaling limits:
$G_L(x/L,0)\propto\text{const.}$ for $x\sim L$, $G_L(x/L,0)\propto
L/x$ for $x\ll L$ and $G_L(0,t/L^{3/2})\propto\text{const.}$ for
$t\gg L^{3/2}$, $G_L(0,t/L^{3/2})\propto L/t^{2/3}$ for $t\ll
L^{3/2}$. For $L\rightarrow\infty$ we obtain
$G_L(x/L,t/L^{3/2})\rightarrow (L/x)G(x/t^{2/3})$ in conformity
with (\ref{scalform}).

It is an important feature of the two-soliton expression
(\ref{cor2}) that the dynamical soliton interpretation directly
implies the correct dependence on the scaling variables $x/L$ and
$t/t_{\text{co}}\propto t/L^{3/2}$, independent of a
renormalization group argument. However, the scaling limits are at
variance with $G_L$. Setting, according to (\ref{cor2}) $\langle
uu\rangle(xt)=(\ell_0/L)F(x/L,t/t_{\text{co}})$, $F(x/L,0)$
assumes the value $.47$ for $x\ll L$ and decreases monotonically
to the value $\sim .08$ for $x\sim L$, whereas $G_L$ diverges as
$L/x$ for $x\ll L$. Likewise, $F(0,t/t_{\text{co}})$ decays from
$.47$ for $t\ll t_{\text{co}}\propto L^{3/2}$ to $0$ for $t\gg
t_{\text{co}}$; for $t\sim t_{\text{co}}$ we have $F_2\sim .15$,
whereas  $G_L$ diverges as $L/t^{2/3}$ for $t\ll t_{\text{co}}$.

This discrepancy from the scaling limits is a feature of the
two-soliton contribution which only samples the correlation from a
single soliton pair. Moreover, at long times the soliton
contribution vanishes and the scaling function is determined by
the diffusive mode contribution in accordance with the convergence
of the phase space orbits to the stationary zero-energy manifold.
We note, however, the general trend towards a divergence for small
values of $x$ and $t$ is a feature of $F$.

Introducing the scaling variables $w=x/\xi\propto x/t^{2/3}$ and
$\tau=t/t_s\propto t/L^{3/2}$ we can also express (\ref{cor2}) in
the form
\begin{eqnarray}
\langle uu\rangle(xt)=(\ell_0/L)F(w,\tau)~,
\end{eqnarray}
where the scaling function $F$ is now given by
\begin{eqnarray}
F(w,\tau)= \frac { \int du e^{-\frac{4}{3}|u|^3\tau}
e^{-4u^2|w\tau^{2/3}+u\tau|} F_1(u) } { \int du
e^{-\frac{4}{3}|u|^3\tau} F_2(u) } ~, \label{cor3}
\end{eqnarray}
and summarize our findings in Fig. \ref{fig15} where we have
depicted $F(w,\tau)$ for a range of $\tau$ values. For fixed small
$w=x/\xi\propto x/t^{2/3}$ we have $F\rightarrow .47$ for
$\tau=t/t_s\propto t/L^{3/2}\rightarrow 0$; for large $\tau$ we
obtain  $F\rightarrow 0$. The weak maximum moving towards smaller
values of $w$ for decreasing $\tau$ is a feature of the functional
form of $F$ in (\ref{cor3}) and thus due to the soliton
approximation. The true scaling function is not expected to have
any particular distinct features
\cite{Hwa91,Tauber95,Frey99,Laessig95,Laessig98a,Laessig00,
Praehofer00a,Colaiori01a,Colaiori01b,Halpin95}.

\section{\label{sum} Summary and Conclusion}
In  the present paper we have continued our analysis of the noisy
Burgers equation in one spatial dimension within the weak noise
canonical phase space approach developed in previous papers. We
believe that the noisy Burgers equation or the equivalent KPZ
equation, which have been studied intensively, is of fundamental
and paradigmatic significance in the context of a continuum field
theoretical description of nonlinear non equilibrium phenomena.
The advantage of the canonical phase space method which is an
elaboration and a dynamical system theory interpretation of the
saddle point equations originating from the Martin-Siggia-Rose
functional formulation or, equivalently, a phase space formulation
of the Freilin-Wentzel variational approach to the Fokker-Planck
equation, actually dating back to work by Machlup and Onsager
\cite{Machlup53,Graham84}, is that it replaces the stochastic
Langevin equation with coupled deterministic field equations,
yielding on the one hand an interpretation of the growth
morphology and pattern formation and on the other hand a practical
scheme for the evaluation of the statistical properties and
correlations in the weak noise limit.

Here we have in some detail discussed i) the growth morphology
engendered by the propagation of domain walls or solitons, the
growth modes, ii) the superimposed linear modes and their
transmutation to propagating modes in the presence of the growth
modes, and, finally, iii) the statical and scaling properties,
particularly, in the two-soliton sector. The weak noise  theory of
the one dimensional Burgers or KPZ equation is, however, far from
complete and many open questions remain. We mention below a series
of topics which it would be of considerable interest to
investigate: i) the interpretation of the solitoninc growth
picture in the context of the mapping of the KPZ equation to the
model of directed polymers in a random medium, ii) a more complete
analysis of the multi-soliton correlations in the thermodynamic
limit with the purpose of making contact with other models in the
KPZ universality class, e.g. the polynuclear growth model
\cite{Praehofer00a,Praehofer00b,Praehofer00c,Praehofer00d}, iii)
elaboration of the anomalous diffusion of growth modes, iv)
contact with other models for many-body systems far from
equilibrium, e.g. driven lattice gas models \cite{Schuetz00}, and,
finally, iv) the extension of the weak noise approach to higher
dimensions.
\begin{acknowledgments}
The author wishes to thank A. Svane for numerous fruitful
discussions. Discussions with J. Hertz and J. Krug are also
gratefully acknowledged. This work has been supported by SNF grant
no. 51-00-0349.
\end{acknowledgments}

\newpage



\begin{figure}
\includegraphics[width=0.6\hsize]
{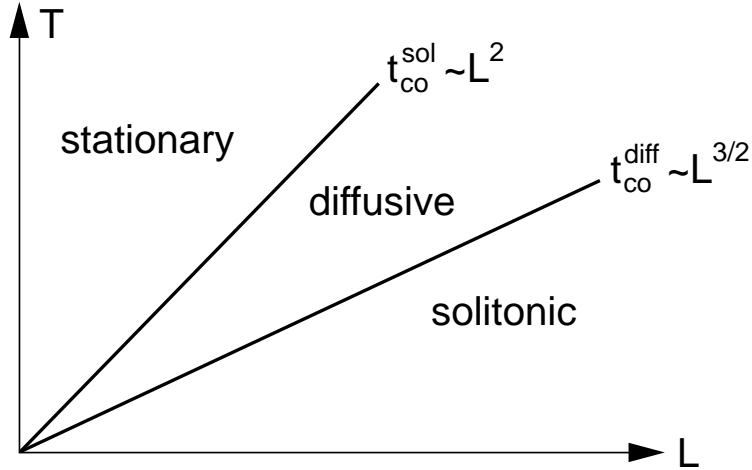} \caption{ In the early time regime for $T\ll
t_{\text{co}}^{\text{sol}}$ the distribution is dominated by
solitons. In the intermediate time regime for
$t_{\text{co}}^{\text{diff}}\gg T\gg  t_{\text{co}}^{\text{sol}}$
the solitons  become suppressed and are replaced by the diffusive
modes. Finally, for $T\gg t_{\text{co}}^{\text{diff}}$ the
diffusive modes also die out and we approach the stationary
distribution. } \label{fig14}
\end{figure}
\begin{figure}
\includegraphics[width=0.6\hsize]
{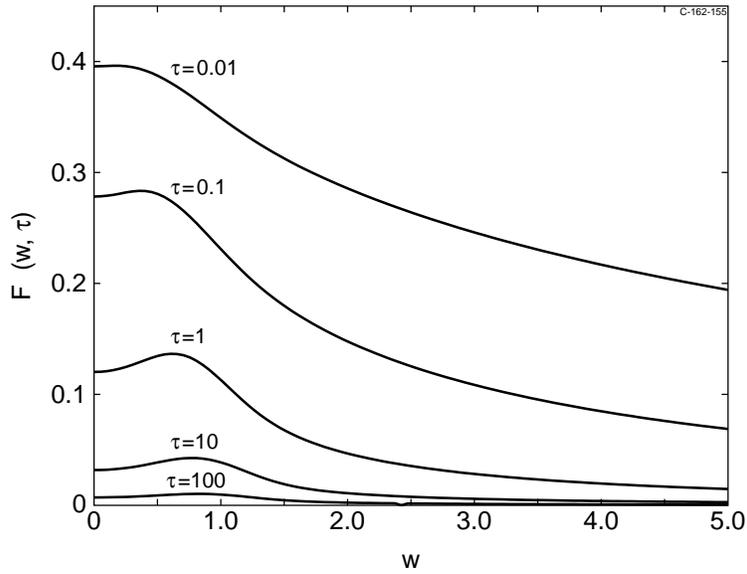} \caption {Plot of the scaling function $F(w,\tau)$
as a function of the scaling variable $w=x/\xi\propto x/t^{2/3}$
for a range of values of $\tau=t/t_{\text{co}}\propto t/L^{3/2}$.}
\label{fig15}
\end{figure}

\end{document}